\documentclass[english]{article}
\usepackage{charter}
\usepackage[T1]{fontenc}
\usepackage[latin9]{inputenc}
\usepackage{geometry}
\usepackage{rotfloat}
\usepackage{amsmath}
\usepackage{amssymb}
\usepackage{graphicx}
\usepackage{setspace}
\makeatletter

\newcommand{\noun}[1]{\textsc{#1}}

\newenvironment{lyxcode}
{\par\begin{list}{}{
\setlength{\rightmargin}{\leftmargin}
\setlength{\listparindent}{0pt}% needed for AMS classes
\raggedright
\setlength{\itemsep}{0pt}
\setlength{\parsep}{0pt}
\normalfont\ttfamily}%
 \item[]}
{\end{list}}

\usepackage{latexsym}\usepackage{graphics}\@ifundefined{definecolor}
 {\usepackage{color}}{}

\makeatother

\usepackage{babel}
\begin{document}

\title{The Kinematics of Completely-Faceted Surfaces}

\author{Scott A. Norris and Stephen J. Watson}

\date{12 October 2009}
\maketitle
\begin{abstract}
We fully generalize a previously-developed computational geometry
tool \cite{norris-watson-2007:threefold} to perform large-scale simulations
of arbitrary two-dimensional faceted surfaces $z=h(x,y)$. Our method
uses a compact, three-component facet / edge / junction storage model,
which by naturally mirroring the intrinsic surface structure allows
both rapid simulation and easy extraction of geometrical statistics.
The bulk of this paper is a comprehensive treatment of topological
events, which are detected and performed explicitly. In addition,
we also give a careful analysis of the subtle pitfalls associated
with time-stepping schemes for systems with topological changes. The
method is demonstrated using a simple facet dynamics on surfaces with
three different symmetries. Appendices detail the reconnection of
{}``holes'' left by facet removal and a strategy for dealing with
the inherent kinematic non-uniqueness displayed by several topological
events. 
\end{abstract}

\section{Introduction}

In many crystal-growing procedures of interest, a nano-scale faceted
surface appears and proceeds to evolve, often exhibiting coarsening
and even dynamic scaling, whereby characteristic statistics describing
the surface remain constant even as the characteristic lengthscale
increases through the vanishing of small facets. For many evolving
faceted surfaces, a \emph{facet velocity law} can be observed \cite{kolmogorov-1940:geometric-selection,drift-1967:diamond-evolution-selection},
assumed \cite{pfeiffer-1985:faceted-silicon,shangguan-hunt-1989:binary-solidification-facet-dynamics},
or derived \cite{gurtin-voorhees-1998,watson-2003:physicaD,watson-2006:prl}
which specifies the normal velocity of each facet, often in configurational
form which depends on the geometry of the facet. In this way, the
dynamics of a continuous, two-dimensional surface can be concisely
represented by a discrete collection of such velocities, and overall
computational complexity reduced to that of a system of ODE's; we
name the resulting system a \emph{Piecewise-Affine Dynamic Surface}
(or PADS). Such theoretical simplification, in turn, enables the large-scale
numerical simulations necessary for the statistical investigation
of coarsening and dynamic scaling.

The numerics involved in the direct geometric simulation of an arbitrary
PADS is straightforward for one-dimensional surfaces, requiring nothing
beyond traditional ODE techniques except simple geometric translation
between facet displacement and edge displacement, and a small surface
correction associated with each coarsening event. Consequently, such
simulations accompany many of the above facet treatments of facet
dynamics, and have also been independently repeated elsewhere \cite{wild-1990:texture-diamond,thijssen-1992:diamond-dynamic-scaling,paritosh-srolovitz-1999:faceted-films,zhang-adams-2002:FACET,zhang-adams-2004:FACET-followup}.
However, in two dimensions, the corrections due to coarsening events
are much more involved, and any code must be able to deal with a family
of non-coarsening \emph{topological events} that alter the neighbor
relations between nearby facets. Consequently, the fewer simulation
attempts use either fast but potentially imprecise envelope methods
\cite{thijssen-1995:diamond-simulation-2,barrat-1996:3D-CVD-diamond,schulze-kohn-1999:geometric-spiral-growth},
or more robust but slower phase-field \cite{taylor-1998:faceted-phase-field,eggleston-2001}
or level-set \cite{russo-smereka-2000:level-set-facets,smereka-2005:level-set-facets,ophus-etal-AM-2009}
methods to avoid explicitly performing topological changes. Besides
the speed/accuracy trade-off exhibited by these approaches, both methods
obscure the natural geometric simplicity of the native surface, complicating
the extraction of detailed surface statistics which, after all, motivates
large simulations in the first place. Additionally, as will be seen,
the presence of non-unique topological events requires explicit intervention
regardless of topological scheme, which negates much of the advantage
of a {}``hands-free'' treatment.

In previous work \cite{norris-watson-2007:threefold}, we introduced
a direct-simulation method which explicitly performs topological events
along the way, thus preserving both simulation speed and topological
accuracy. In addition, by representing the surface as a collection
of facets, edges, and junctions, plus the neighbor relations between
them, the method mirrors the natural geometry of the surface being
modeled, which allows easy extraction of geometric statistics. There,
however, the restricted case of threefold symmetry was chosen for
ease of topological implementation; under this symmetry, a limited
number of topological events were observed, and both vanishing facets
and non-vanishing surface rearrangements could be handled explicitly
using a priori knowledge of the before and after surface states. While
many surfaces exhibit threefold symmetry, making the method useful
even in this special case, it could not handle other common crystal
symmetries, notably fourfold and sixfold.

Here, we generalize the previous model to allow the simulation of
surfaces with arbitrary symmetry groups. We begin in Section~\ref{sec: cellular-structure}
with a brief summary of the basic method, including surface representation,
facet kinematics, and the application of a dynamics. Next, in Section~\ref{sec: topological events},
we provide a careful enumeration of topological events which may occur
on surfaces of arbitrary symmetry; this includes discussion of the
Far-Field Reconnection algorithm (FFR), by which network holes left
by vanishing facets may be consistently repaired without knowledge
of the post-event state. Then, we provide in Section~\ref{sec: discretization and topology}
a careful consideration of the consequences of using (necessarily
discrete) timesteps during the simulation of a surface whose evolution
equations change qualitatively between steps (at topological events);
the issues that arise are discussed in the context of three sample
strategies. The completed method is illustrated from three-, four-,
and six-fold symmetric surfaces in Section~\ref{sec: demonstration};
these exhibit all of the topological events likely to be encountered
on a real surface, and demonstrate that the method is robust enough
to generically simulate faceted surfaces of any symmetry class for
which a facet-velocity law is uniquely specified. Finally, in addition
to detailing the FFR algorithm, the appendix includes a discussion
of kinematically non-unique topological events, where two resolutions
are possible, and highlights the need to refer to the dynamics or
even first principles to decide how the surface should evolve in those
cases.

\section{Data structures and simple motion: a 3D cellular network\label{sec: cellular-structure}}

\subsection{Characterization\label{sec: characterization} }

\begin{figure}[ht]

\begin{centering}
\scalebox{.5}{\includegraphics{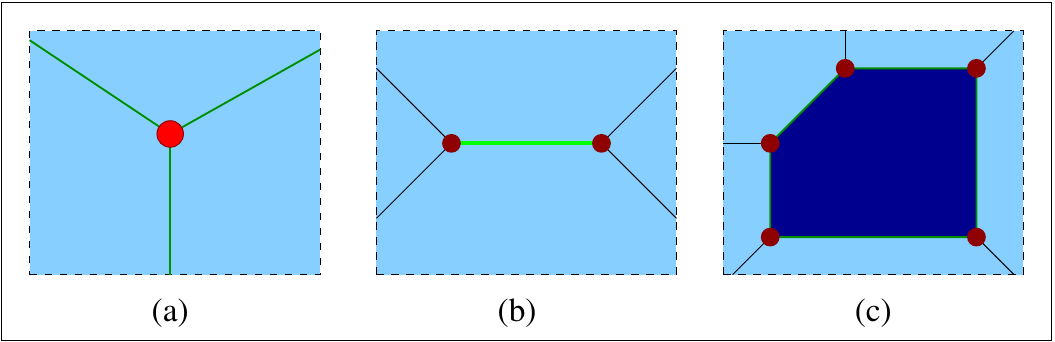}}
\caption[Diagram of neighbor relations]{ Neighbor relations for each kind of surface element. (a) A junction
neighbors three edges and three faces. (b) An edge neighbors two junctions
and two faces. (c) A facet neighbors $m$ junctions and $m$ edges
(here $m=5$). }

\par\end{centering}

\label{fig: neighbor-relations} 
\end{figure}

We consider the evolution of a single-valued, fully-faceted surface
$z=h(x,y,t)$; this definition explicitly forbids overhangs and inclusions.
We assume that the surface bounds a single crystal which exists on
exactly one lattice; thus, we are not treating surfaces with multiple
grains. The surface is piecewise-affine, consisting of facets $\{\mathcal{F}_{i}\}$
with fixed normals $\{\boldsymbol{n}_{i}\}$. These are bounded by
and meet at edges $\mathcal{E}$ which are necessarily straight line
segments; edges in turn meet at triple-junctions $\mathcal{J}$. This
three-component structure is reminiscent of two-dimensional \emph{cellular
networks} \cite{glazier-weaire-1992,stavans-1993,frost-thompson-1996,thompson-2001,schliecker-2002}
and indeed, while we consider three dimensional surfaces, the projection
of the edge set onto the plane $z=0$ is a 2D cellular network. This
structure and the neighbor relations inherent within it suggest a
doubly-linked object-oriented data structure, consisting of: (1) a
set of junctions, each having a location, pointing to three edges
and three facets; (2) a set of edges, each having a tangent, pointing
to two junctions and two facets; and (3) a set of facets, each having
a normal, pointing to $m$ edges and $m$ junctions. These objects
and the associated neighbor relations are illustrated in Figure~\ref{fig: neighbor-relations};
this structure is the natural structure of the surface, and uniquely
and exactly describes it. We now consider each element in more detail.

\subsubsection{Junctions\label{sec: junctions}}

A \emph{junction} is a point in space formed where edges (and hence,
facets) intersect. The \textbf{order} $n$ of a junction is simply
the number of edges which meet there. While junctions of any order
$n\ge3$ are possible, we restrict ourselves here to the case of order
$3$ junctions or {}``triple junctions.'' This greatly simplifies
analysis and code, as triple junctions are uniquely positioned by
the three facets meeting there. The intrinsic geometric information
carried by a junction is its location. Junctions are stored in a \noun{Junction}
class, which contains this location, as well as pointers to the three
edges and three facets which meet there.

\subsubsection{Edges\label{sec: edges}}

An \emph{edge} is a line segment formed by the intersection of exactly
two facets, and bounded by exactly two junctions. The intrinsic geometrical
quantity of an edge is its orientation, which is fixed since facets
have fixed normals. Edges are stored in the \noun{Edge} class, which
records the tangent, as well as pointers to the two neighboring facets
and two bounding junctions. At creation, edges are {}``directed'':
one junction is arbitrarily deemed the origin, and the other the terminus,
establishing a \emph{tangent}. This has two important consequences.
First, if we imagine walking along the edge in the tangent direction,
then one neighboring facet may be labeled {}``left'', and the other
{}``right.'' This information allows us to distinguish between convex
and concave edges, and also to determine the clockwise direction around
a given facet, which is necessary for effective navigation of the
network, as well as the proper calculation of boundary integrals on
facets. Second, the tangent allows us to detect topology changes \emph{a
posteriori}, when an edge {}``flips'' (see \cite{roosen-taylor-1994:faceted-interface});
this will be discussed in more detail in Section~\ref{sec: topology-class-1}.

\subsubsection{Facets\label{sec: faces}}

A facet is a simply-connected planar polygonal region in space, which
is bounded by an equal number of edges and junctions. The intrinsic
geometric information carried by a facet is its normal, which is fixed.
Our surface definition $z=h(\boldsymbol{x},t)$ requires that the
normal of each facet is constrained to be on the hemispherical shell
of unit-length vectors with positive $z$ component. The imposition
of a particular symmetry on the crystal may further restrict available
normals, but no such restriction is here assumed. Facets are stored
in a \noun{Facet} class, which contains the normal, as well as a list
of bounding edges and junctions, sorted in counter-clockwise order
(which is enabled by the directedness of the edges). From this stored
information, and the information stored in the junctions and edges
bounding a facet, it is a simple task of vector calculus to calculate
geometric quantities such as the perimeter, area, and center of mass
of the facet.

\subsection{Kinematics}

The intrinsic geometric means of characterizing surface evolution
is by specifying the normal velocity of each point on the surface.
A piecewise-affine surface is composed of a collection of planar,
fixed-normal facets, whose motion is limited to displacement along
the normal. Therefore, the kinematics $V_{n}$ of the entire surface
may be expressed by a discrete set of individual facet velocities
$V_{i}$. As edges and junctions are merely intersections between
two and three facets, respectively, their motion is uniquely specified
by the motion of the facets that neighbor them. In particular, if
$\boldsymbol{p}$ is the location in space of a triple junction, then
the velocity of that a triple junction may be calculated through the
expression 
\begin{equation}
\frac{d\boldsymbol{p}}{dt}=\boldsymbol{A}^{-1}\boldsymbol{v},
\end{equation}
where the rows of $\boldsymbol{A}$ and entries of $\boldsymbol{v}$
are the unit normals and normal velocity, respectively, of the three
facets intersecting to form $\boldsymbol{p}$. In practice, facet
velocities are only used indirectly to calculate junction velocities
-- if junctions are moved correctly, edges (connections between two
junctions) and thus faces (collections of edges) are necessarily moved
correctly as well.

\subsection{Dynamics}

All that remains now is to select a particular dynamics; that is,
to specify an expression for the normal velocity $V_{i}$ of each
facet. Having chosen one, we follow \cite{watson-2006:prl} and refer
to the resulting evolving structure as a Piecewise-Affine Dynamic
Surface (or PADS). Example dynamics describing many different physical
situations were listed in the introduction, and the exact dynamics
is not of special concern here (although we will select one for demonstration
later). It is worth noting here, however, that most of the dynamics
proposed to date are \emph{configurational}, depending on properties
of the facet such as area, perimeter, number of junctions, or mean
height. Thus, sudden changes in the geometric properties of a facet
can lead to sudden changes in its velocity, an issue which will be
explored in more detail in Section~\ref{sec: discretization and topology}.

\section{Topological Events\label{sec: topological events}}

We have just discussed how elements of each class (facet, edge, vertex)
neighbor members from each of the other classes. Taken together, the
set of all of these neighbor relations comprises the \textbf{topological
state} of the surface. It is a complete record of every neighbor relationship
on the surface, and is unique for a given surface. As the system evolves,
these neighbor-relations may change as facets exchange neighbors,
join together, split apart, or vanish. Each of these cases is an example
of a \textbf{topological event}, and represents a change to the topological
state of the surface (topological events are a defining feature of
evolving cellular networks -- again see \cite{glazier-weaire-1992,stavans-1993,frost-thompson-1996,thompson-2001,schliecker-2002}).
To maintain an accurate representation of the surface, a direct geometric
method like that described here must manually perform topological
events as necessary. Because actual surface evolution is fairly trivial,
this is the main difficulty of our method.

A natural first question to ask at this point is {}``how many topological
events are possible?'' To begin answering this question, we point
out that on a physical surface, topological events occur automatically,
and by geometric necessity. If a detected event signals the need to
change neighbor relationships at some location on the surface, we
may therefore infer that failing to change them would produce a cellular
network with {}``wrong'' relationships, that do not correspond to
a physical surface. We call such erroneous configurations \textbf{geometrically
inconsistent}; examples include primarily edge networks that intersect
when viewed from above, since these correspond to overhangs and inclusions,
which are prohibited. Since topological events serve to avoid possible
geometric inconsistencies, we may discover what events are possible
by considering how inconsistencies may occur. This is most easily
accomplished by considering each surface element in turn.

We first consider junctions, which are simply a location in space.
A junction can, in the course of surface evolution, leave the periodic
domain, in which case it is wrapped to the other side. However, this
is only a bookkeeping operation, and does not represent a real topological
event. Turning to edges, we note that edges possess a directed length.
As already hinted in section~\ref{sec: edges}, this length could
become negative if the edge were to {}``flip'' \cite{roosen-taylor-1994:faceted-interface}
A flipped edge has no geometrical meaning on a single-valued surface,
and so we introduce a class of \textbf{Vanishing Edge} events which
occur when edges reach zero length. Finally, we consider facets. Since
a facet has fixed orientation, its changing properties are loosely
its shape and size. Specifically, a facet is a \emph{simply connected}
planar region with \emph{positive area}. These two defining properties
of facets lead, through consideration of their potential violation,
to two additional classes of topological event: \textbf{Facet Constriction}
events which prevent the formation of self-intersecting facets, and
\textbf{Vanishing Facet} events which remove facets from the network
when they reach zero area.

\subsection{Vanishing Edges\label{sec: topology-class-1}}

An vanishing edge event occurs when an edge shrinks to zero length,
and its junctions meet. To consider what might happen to the faceted
surface when this occurs, we first label the immediate surroundings
of an edge. Each edge is composed of two faces of which it is the
intersection, its \textbf{composite faces}, and stretches between
two faces at which it terminates, its \textbf{terminal faces}. In
addition, we will also use the term \textbf{emanating edges} to refer
to those edges immediately neighboring the shrinking edge. Now, consider
the hemispherical shell of available facet normals (Section~\ref{sec: faces}).
The (necessarily distinct) normals of the composite faces specify
a great circle about this hemisphere, which divides it into two parts.
The normals of the terminal faces cannot lie on this boundary - otherwise
their intersection with the composite faces would form lines instead
of junctions. Furthermore, unless they are identical (a special case),
they form a second great circle around the hemisphere. While terminal
normals may not lie on the composite great circle, the reverse is
not true, and this fact effectively divides Vanishing Edge events
into three sub-classes.

\begin{figure}[ht]

\begin{centering}
\scalebox{.5}{\includegraphics{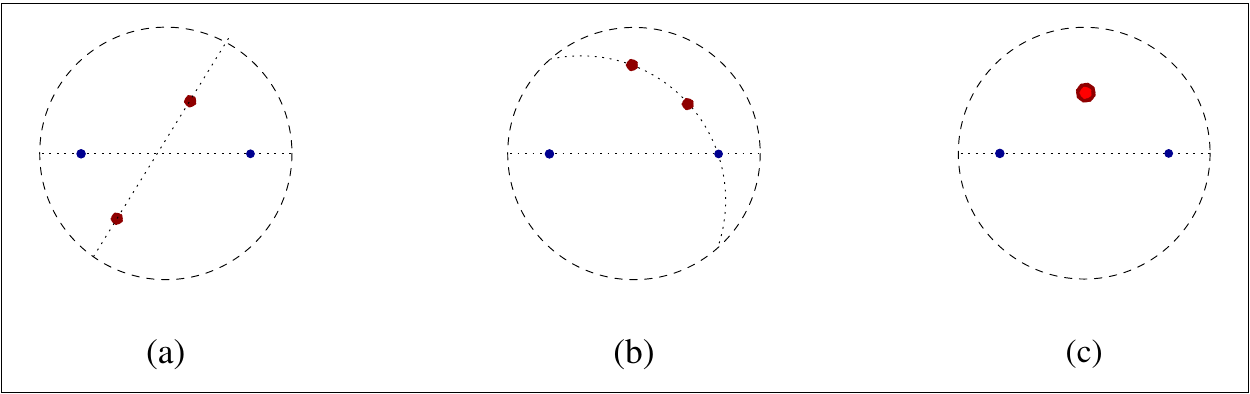}}
\caption[Normal diagrams for Vanishing Edge events]{ Normal Diagrams for different types of Vanishing Edge events. Blue
dots represent the normals of composite faces, while red dots represent
the normals of terminal faces. Dotted lines represent great circles
between two points. (a) Terminal great circle touches neither composite
point. (b) Terminal great circle touches one composite point. (c)
Terminal normals occupy the same point. Great circle undefined. }

\par\end{centering}

\label{fig: shrinking edge classes} 
\end{figure}

Figure~\ref{fig: shrinking edge classes} illustrates this idea,
and gives an example of each of the three possible cases. If the terminal
great circle touches neither composite point, then the well-studied
\textbf{Neighbor Switch} occurs. If the terminal great circle touches
one composite point, then an \textbf{Irregular Neighbor Switch} results.
Finally, if the terminal normals occupy the same point, then no great
circle is defined -- the terminal facets have he same normal, and
when the edge between them shrinks to zero, they join into a single
facet: a \textbf{Facet Join}.

\subsubsection{Neighbor Switch}

\label{sec: t1}

\begin{figure}[ht]

\begin{centering}
\scalebox{.5}{\includegraphics{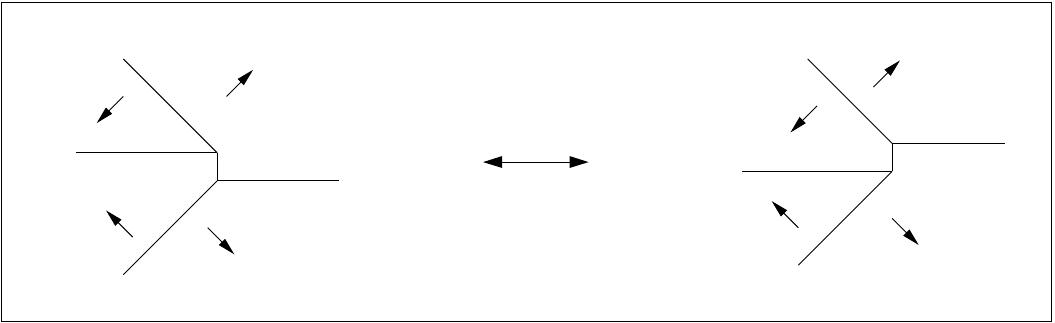}}
\\
 %\scalebox{.5}{\includegraphics{figures/illustrations/general/t1-schematic.eps}}
 \caption[Neighbor Switch event]{ Example of a Neighbor Switch Event. Arrows represent gradients of
regions in which they appear. }

\par\end{centering}

\label{fig: t1-illustration} 
\end{figure}

On a general surface, the most common Vanishing Edge event is the
neighbor switch, which is frequently encountered in other evolving
cellular networks. In this event, neither composite normal touches
the terminal great circle, so any three of the normals involved form
a linearly independent set -- this property is the defining feature
of the neighbor switch. When an edge with this configuration shrinks
to zero length, the surrounding facets simply exchange neighbors.
Figure~\ref{fig: t1-illustration} gives an example of this event.

\emph{Resolution.}\,\, To resolve this event, we simply delete the
old edge, and create a new edge. The composite faces which formed
the old edge become terminal faces of the new edge, and cease to neighbor
each other. Conversely, the terminal faces of the old edge become
the composite faces of the new edge, and thus become neighbors. This
symmetric exchange in neighbor relations is the cause of the name
Neighbor Switch, which comes from the grain-growth literature -- the
less-descriptive name {}``T1 process'' in often used in the soap
froth literature. In addition to replacing the vanishing edge, the
junctions on either side of this edge are replaced. Each new junction
is formed by the intersection of the deleted edge's (formerly non-adjacent)
terminal faces with one of its composite faces.

\emph{Comments.}\,\, Readers familiar with other cellular-network
literature will note that the example Neighbor Switch in Figure~\ref{fig: t1-illustration}
lacks the typical {}``X'' shape. This is due to the constrained
nature of facet normals, and hence, edge orientations. Additionally,
we note that the neighbor switch is a reversible event; in fact it
is its own reversal. Finally, a certain sub-class of neighbor switches
possessing {}``saddle'' structure are non-unique, as was observed
by Thijssen \cite{thijssen-1995:diamond-simulation-2}. For a discussion
of this non-uniqueness and its consequences, see Appendix~\ref{sec: non-uniqueness}.

\subsubsection{Irregular Neighbor Switch}

\label{sec: gap-opener}

\begin{figure}[ht]

\begin{centering}
\scalebox{.5}{\includegraphics{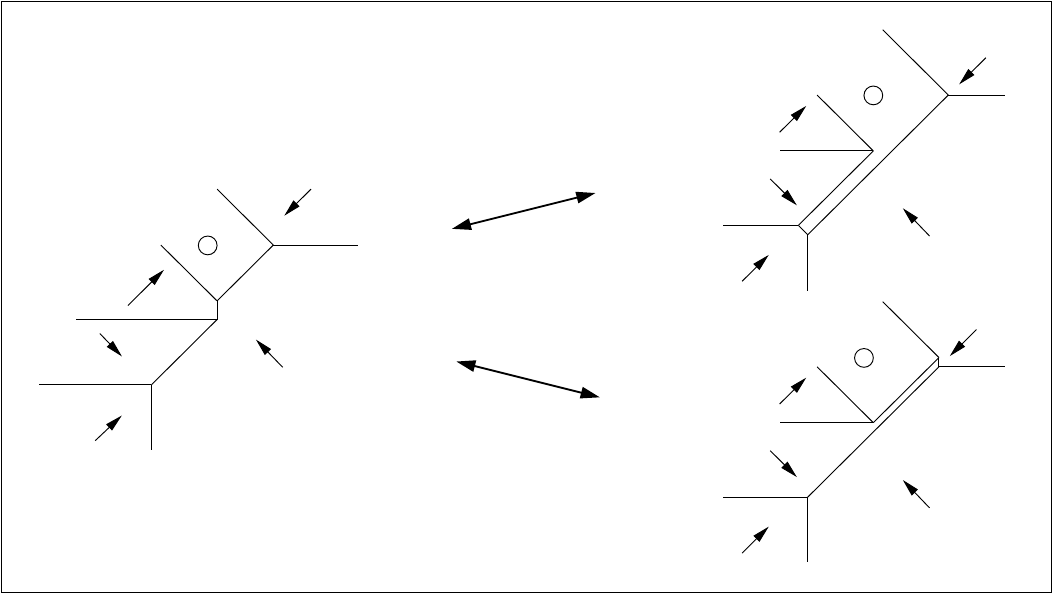}}
\caption[Irregular Neighbor Switch event]{ An example of the Irregular Neighbor Switch event. From left to
right is the non-unique {}``gap opener.'' From right to left is
the unique {}``gap closer.'' Arrows represent gradients of regions
in which they appear, while circles indicate flat facets with zero
gradient (vertical normal). }

\par\end{centering}

\label{fig: gap opener} 
\end{figure}

When the normal of one of the composite faces lies on the great circle
formed by the terminal normals, the neighbor switch cannot occur.
Here, the terminal faces cannot form a new junction with the offending
composite face because the three normals involved are not independent.
Instead, when an edge with this configuration shrinks to zero, two
closely related events are possible, depending on the configuration
of the nearby edges. These events are collectively called Irregular
Neighbor Switches, with two varieties called a {}``gap opener''
and {}``gap closer'' that are exact opposites. These are illustrated
in figure~\ref{fig: gap opener}.

\emph{Resolution.}\,\, Because one composite normal lies on the
terminal great circle, exactly two of the emanating edges are parallel
in $\mathbb{R}^{3}$. The gap opener occurs when these edges emanate
from the shrinking edge in \emph{opposite} directions, while the gap
closer occurs when the edges emanate in the \emph{same} direction.
To resolve the gap opener, we select one of the parallel emanating
edges to be split apart (see below). The gap will go here, filled
by the terminal face that touches the other parallel edge, and will
extend all the way to the far end of the split edge, where a new edge
is introduced to link the two edges resulting from the split edge.
This is all illustrated in Figure~\ref{fig: gap opener}. To resolve
the gap closer, simply reverse the steps.

\emph{Comments.}\,\, Several comments on this pair of events are
in order. First, while the gap closer is uniquely resolved, the gap-opener
is an inherently non-unique event, as either of the parallel edges
could be the one split (we will discuss this further in section~\ref{sec: non-uniqueness}).
Second, both resolution options have the potentially dissatisfying
property of being non-local in effect, because the collision of two
junctions causes an entire edge to split apart. What is perhaps more
likely is the nucleation of a new, tiny facet at the moment the junctions
collide; however, we have not considered the case of nucleation in
this work. Finally, while common experimentally-encountered surfaces
usually have either high symmetry (only a few facet orientations)
or no symmetry (as many orientations as facets), the irregular neighbor
switch with its three coplanar orientations requires what may be called
{}``intermediate symmetry,'' where a large, but still limited, number
of orientations are available. Hence, this event may likely never
be encountered in a physically-relevant surface. We include it here
for completeness.

\subsubsection{Facet Join}

\label{sec: face-join}

\begin{figure}[ht]

\begin{centering}
\scalebox{.5}{\includegraphics{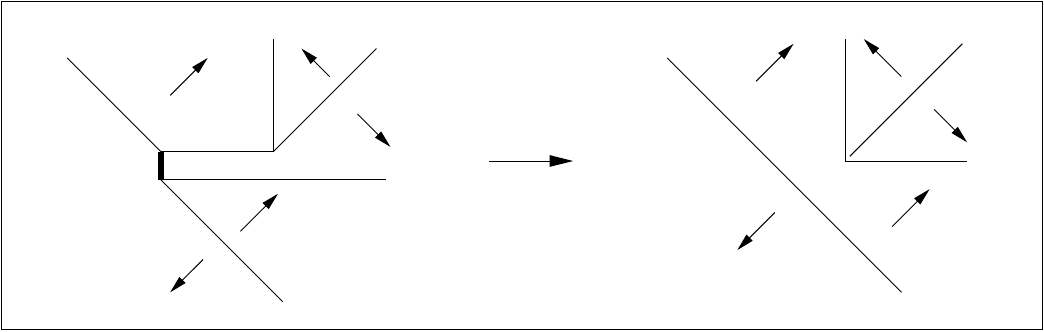}}
%\scalebox{.5}{\includegraphics{figures/illustrations/general/fjoin-schematic.eps}}
 \caption[Facet Join event]{ An example facet join procedure. Arrows represent gradients of regions
in which they appear. %(b) schematic of a face-join procedure.
 %Dotted regions have the same orientation.
 }

\par\end{centering}

\label{fig: face-join-illustration} 
\end{figure}

Finally, we consider the special case where the terminal normals are
identical. When such an edge shrinks to zero length, the terminal
faces meet exactly. Having the same orientation, they then join to
form a larger face. Figure~\ref{fig: face-join-illustration} depicts
a representative facet join event.

\emph{Resolution.}\,\, Facet Joins are performed by creating a new
facet to replace the joining facets, and all edges and junctions that
neighbored the old facets are re-assigned to this new facet. Next,
the vanishing edge and its two junctions are deleted, leaving the
four emanating edges to be considered. These are most logically grouped
into the (necessarily parallel) pairs of edges bordering, respectively,
the left and right composite faces of the vanished edge. In the example
event shown in Figure~\ref{fig: face-join-illustration}, these two
pairs look different: one pair meets side-to-side, while the other
pair meets end-to-end. Computationally, however, this makes no difference;
each pair is replaced by a single edge connecting their remaining
non-deleted junctions. This behavior is generic for all facet joins.

\emph{Comments.}\,\, We note that the face join is, strictly speaking,
non-reversible (though see Section~\ref{sec: face-split}). The exact
opposite of the face join would be a facet which spontaneously {}``shatters,''
as described in \cite{roosen-taylor-1994:faceted-interface}; this
behavior is certainly worth studying, but is not currently implemented.
Second, although this is a {}``special case'' in general, for high-symmetry
crystal surfaces it may be very common -- indeed, for the case of
a cubic crystal with only three available facet orientations, Facet
Joins are the only Vanishing Edge event exhibited \cite{norris-watson-2007:threefold}.
Finally, we note that this event is the only Vanishing Edge event
which does not conserve the number of facets. It is, in fact, one
mechanism by which coarsening may occur, and may be the dominant mechanism
for high-symmetry surfaces \cite{norris-watson-2007:threefold}.

\subsection{Facet Constrictions}

\label{sec: topology-class-2} 
\begin{figure}[ht]

\begin{centering}
\scalebox{.5}{\includegraphics{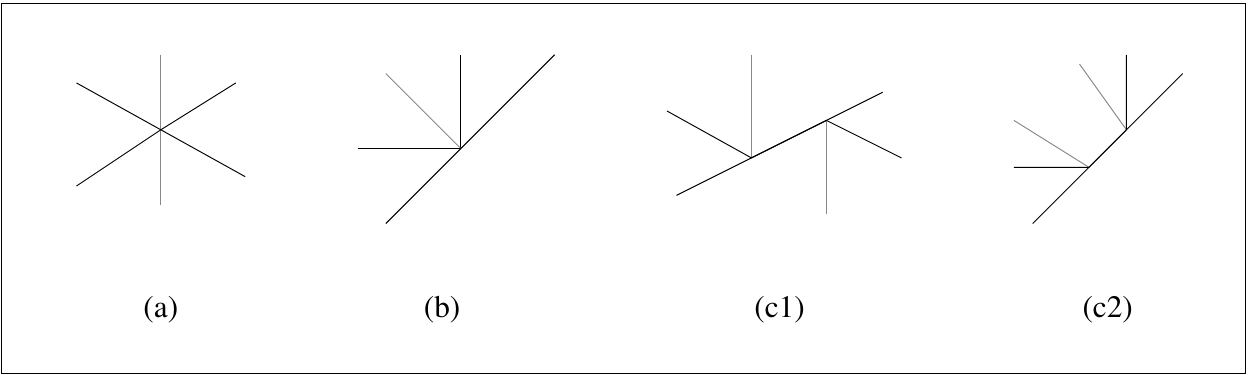}}
\caption[Signatures of Constricted Facet events]{ Signatures of Constricted Facet events. (a) Non-Adjacent Junction-Junction
collision signature. (b) Junction-Edge collision signature. (c1),(c2)
Asymmetric and Symmetric Edge-Edge collision signatures. }

\par\end{centering}

\label{fig: self-intersection-signatures} 
\end{figure}

The second class of topological event
occurs whenever a facet ceases to be simply-connected, and results
in that facet being split into two new facets. Remembering that the
edges of a facet trace out a polygon in the plane, we observe that
the non-simply connected polygon, if allowed to continue evolving,
would become self-intersecting, which clearly has no geometrical interpretation.
So, how may an evolving polygon become self-intersecting? Since the
boundary consists of edges and junctions, there are three possible
modes: (a) two non-adjacent junctions meet, (b) a junction meets an
edge, or (c) two edges meet. Each case has a distinct {}``signature,''
illustrated in Figure~\ref{fig: self-intersection-signatures}, which
can be used to tell them apart.

The Junction-Junction collision shown in Figure~\ref{fig: self-intersection-signatures}a
represents the formation of a perfect $O(6)$ junction. While theoretically
interesting, such events are not considered here; we hypothesize that,
given random initial data, two junctions not connected by an edge
will never exactly meet. Furthermore, by considering Figure~\ref{fig: self-intersection-signatures},
it can be seen that all Junction-Junction collisions, if perturbed
as we hypothesize, result in either junction-edge or edge-edge collisions,
and can therefore be resolved accordingly.

Junction-Edge collisions occur when a facet is pinched into two pieces
by three of its neighbors, depicted in Figure~\ref{fig: self-intersection-signatures}b.
There, two adjacent neighbors of the facet, forming a wedge, meet
a third neighbor and pierce it. Two separate events are possible in
this class. In most cases, the wedge simply splits the central facet
into two parts, in an event called a \textbf{Facet Pierce}. However,
if the normals of the wedge facets and the normal of the central facet
lie on the same great circle, then, as the central facet is split,
the opposing facet opens up a gap in the wedge: an \textbf{Irregular
Facet Pierce}.

Edge-edge collisions occur when a facet is pinched by four neighbors,
shown in Figure~\ref{fig: self-intersection-signatures}c. In these
events, two non-adjacent, exactly parallel edges meet, which requires
that the normals of the impinging facets be coplanar with the normal
of the pinched facet. Again, two variations are possible. If the impinging
faces have different normals, the event is called a \textbf{Facet
Pinch}. However, if they have the same normal, they join even as they
pinch the facet in question, in a process called a \textbf{Joining
Facet Pinch}. In addition, each event may occur in either symmetrical
or asymmetrical flavors, which are shown in Figure~\ref{fig: self-intersection-signatures}c1,c2
respectively. The meeting of two edges requires the involvement of
two junctions; these lie on the same edge for the symmetrical case,
and on different edges for the asymmetrical case, as seen in the figure.

\subsubsection{Facet Pierce}

\label{sec: face-split} 
\begin{figure}[ht]

\begin{centering}
\scalebox{.5}{\includegraphics{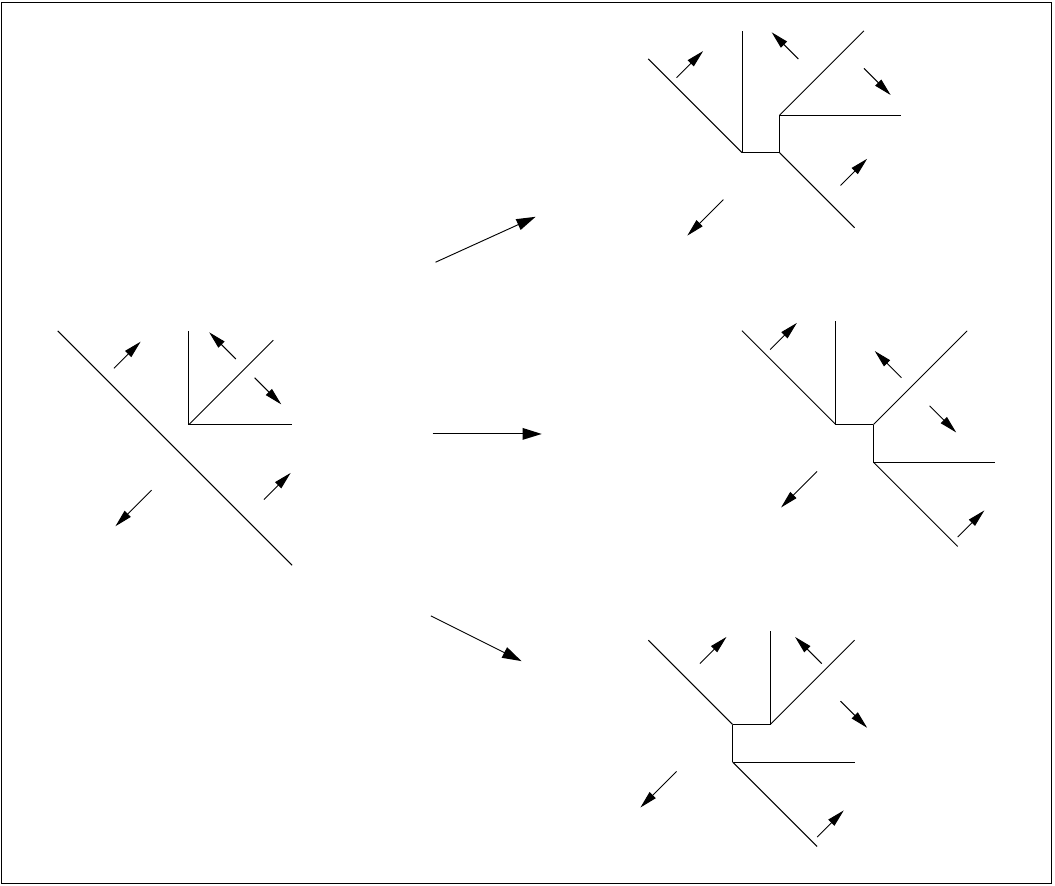}}
%\scalebox{.5}{\includegraphics{figures/illustrations/general/point-line-schematic.eps}}
 \caption[Facet Pierce event]{ Example of a facet pierce procedure. At the moment the event occurs,
an $O(5)$ junction is formed, which immediately breaks in one of
three ways, depending on the dynamics. Arrows represent gradients
of the regions in which they appear. }

\par\end{centering}

\label{fig: face-split-illustration} 
\end{figure}

The first self-intersection we will study
is the simplest; the facet pierce. It is a point-line event as described
above; that is, a facet is split when a triangular wedge formed by
two adjacent neighboring facets intersects the edge formed with a
third, opposing neighbor. The facet pierce is functionally the opposite
of a facet join, and is illustrated in Figure~\ref{fig: face-split-illustration}.

\emph{Resolution.}\,\, Given the constricted facet, as well as the
junction and edge which meet, we can label all of the surrounding
facet elements and deterministically reconnect them correctly. First,
two new facets are created to replace the constricted facet. The junctions
and edges that bordered the old facet can be reassigned to these based
on the labels created initially. The colliding junction and edge are
deleted, to be replaced by three new junctions and two new edges.
The locations of the former and neighbor relationships of each can
be determined by considering Figure~\ref{fig: face-split-illustration}
and using the labels.

\emph{Comments.}\,\, First, technically, at the moment of the event,
an $O(5)$ junction forms, which as shown in Figure~\ref{fig: face-split-illustration}
may proceed to break in one of three ways. This does not, however,
constitute a non-uniqueness; rather, the dynamics governing the surface
evolution at the moment of topological change specify which exit pathway
is chosen. Second, while Thijssen \cite{thijssen-1995:diamond-simulation-2}
rightly objected to this resolution for the case of separate grains,
we find it satisfactory for the case of a single crystal considered
here.

\subsubsection{Irregular Facet Pierce}

\label{sec: table-split} 
\begin{figure}[ht]

\begin{centering}
\scalebox{.5}{\includegraphics{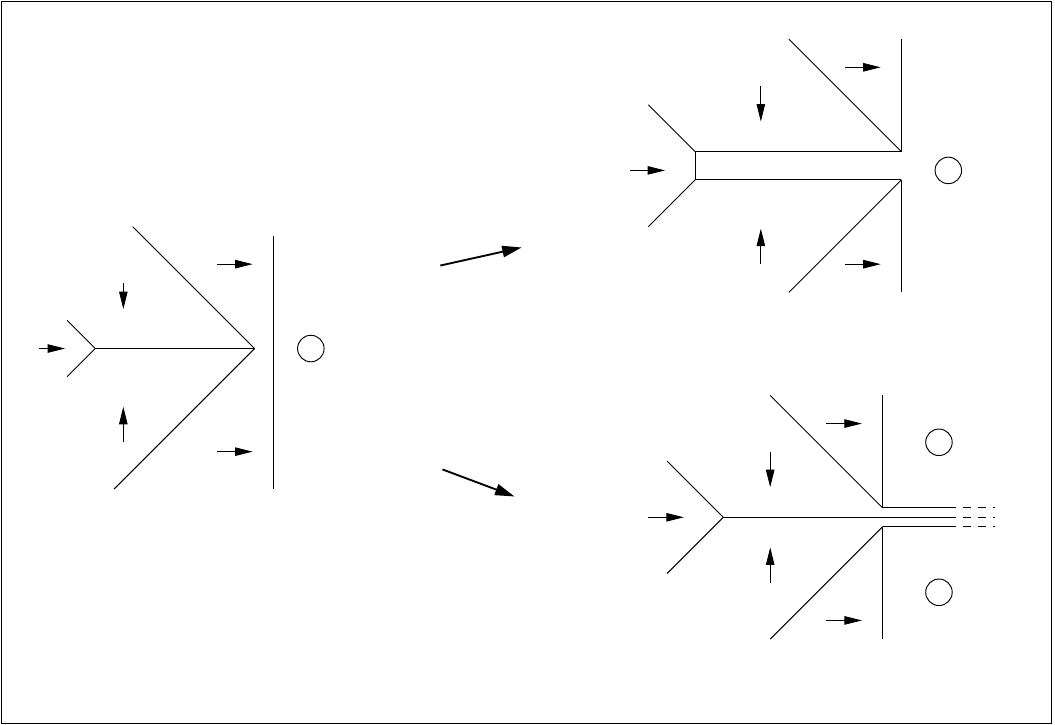}}
\caption[Irregular Facet Pierce event]{ Example of an irregular facet pierce procedure. Arrows indicate
gradients; circles flat planes with no gradient. }

\par\end{centering}

\label{fig: table-split-illustration} 
\end{figure}

A special modification of the Facet Pierce
just described occurs when the normal of the opposing facet shares
a great circle with the normals of the facets forming the wedge. This
event is called an Irregular Facet Pierce. Recall that three new junctions
were created during the facet pierce. However here, since the two
newly created facets have identical normals, and the remaining three
have normals which are not independent, those junctions cannot be
created. Instead, as the wedge facets meet the opposing facet, one
of two things happen -- either the center edge of the wedge is split
apart by the opposing facet (a {}``wedge split''), or the opposing
facet is split apart by the wedge (a {}``wedge extension''). We
see an illustration of each possibility in Figure~\ref{fig: table-split-illustration}.

\emph{Resolution.}\,\, This event is repaired quite similarly to
the regular facet pierce, with modifications. As is done there, two
new facets are created to replace the constricted facet, and junctions
and edges bordering the old facet are reassigned to the new ones.
The resolution differs in how to replace the colliding junction and
edge. If the {}``wedge split''resolution is chosen, then the middle
edge of the wedge and its far junction are also deleted -- these are
replaced by two parallel edges and junctions. Finally, an edge is
formed which links them and borders the facet on the far side of the
deleted edge. If the {}``wedge extension'' resolution is chosen,
not only is the constricted facet split apart, but so is the one opposite
the edge split by the wedge. One must first determine which edge of
this second split facet the extended wedge will intersect. Having
done so, that facet is deleted, to be replaced by two new facets.
The extension is formed by adding two edges parallel to the middle
edge of the wedge, and the edge it intersects is split in two. Two
new edges and three junctions must be created to link the extension
with the edge it intersects. Finally, all edges and junctions bordering
the deleted facet, plus those created to form the extension, are re-assigned
appropriately to the new facets. Figure~\ref{fig: table-split-illustration}
is especially helpful here.

\emph{Comments.}\,\, The event clearly recalls the {}``gap opener''
described above. It shares with that event three coplanar surface
normals, and as a result, two possible resolutions. Additionally,
while the two options here are qualitatively different compared to
the symmetric options of the gap opener, they are additionally both
non-local effects due to a local cause. Again, perhaps the best resolution
is to nucleate a new facet, which we do not yet consider. Finally,
both events require {}``intermediate symmetry,'' and for the same
reasons discussed above, we have not implemented this event.

\subsubsection{Facet Pinch}

\label{sec: face-pinch} 
\begin{figure}[ht]
 %\scalebox{.5}{\includegraphics{figures/illustrations/general/fpinch-schematic.eps}}

\begin{centering}
\scalebox{.5}{\includegraphics{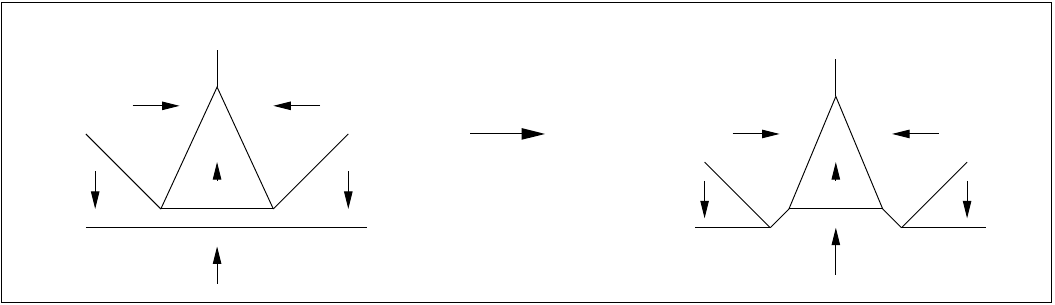}}
\caption[Facet Pinch event]{ %Schematic of half of a face-pinch procedure.
 Example of a symmetric face pinch. Arrows represent gradients of
regions in which they appear. }

\par\end{centering}

\label{fig: face-pinch-illustration} 
\end{figure}

We now turn to consider the case of Edge-Edge
events, the first of which is called a Face Pinch. Here the normals
of the pinching facets are not identical, and so junctions can be
created as needed -- an illustration of this event is shown in Figure~\ref{fig: face-pinch-illustration}.
This event is practically similar to the face split described above.
In each case, a facet is split into two by non-joining neighbors;
the difference is just whether the procedure is {}``sharp'' or {}``blunt'';
i.e., caused by parallel edges or a junction and an edge.

\emph{Resolution.}\,\, Because of the similarities between the facet
pierce and facet pinch, the associated resolutions are similar as
well. Here, with knowledge of the constricted facet and the two colliding
edges, we can label all of the surrounding facet elements and deterministically
achieve the change shown in Figure~\ref{fig: face-pinch-illustration}.
As with the Facet Pierce, two new facets are created to replace the
constricted facet, and the junctions and edges that bordered the old
facet are reassigned as required. The colliding edges are deleted,
as are the associated junctions discussed above. Five edges and four
junctions are created to complete the reconnection, as shown in Figure~\ref{fig: face-pinch-illustration}.

\subsubsection{Joining Facet Pinch}

\label{sec: face-swap} 
\begin{figure}[ht]
 %\scalebox{.5}{\includegraphics{figures/illustrations/general/fswap-schematic.eps}}

\begin{centering}
\scalebox{.5}{\includegraphics{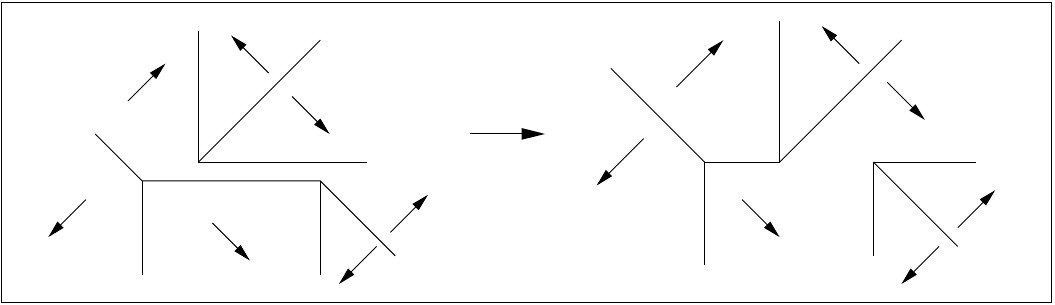}}
\caption[Joining Facet Pinch event]{ %Above: schematic of half of a face-swap procedure.
 %Dotted regions have the same orientation.
 Example of an asymmetric face swap. Arrows represent gradients of
regions in which they appear. }

\par\end{centering}

\label{fig: face-swap-illustration} 
\end{figure}

Finally, a special modification of the face-pinch
occurs when the impinging facets have identical normals. The constricted
facet is split in exactly the same way as in a face pinch; however,
since the two facets doing the {}``pinching'' are identically oriented,
they join together to form a larger facet. We see an illustration
of this situation in Figure~\ref{fig: face-swap-illustration}.

\emph{Resolution.}\,\, Resolutions is similar to that for the Facet
Pierce and Facet Pinch. We again need to know the constricted facet
and the two meeting edges, which allows the necessary labeling. Again,
two new facets are created to replace the constricted facet, but in
this case the two facets which meet must join, and so another new
facet must be created to replace them -- necessary junction and edge
reassignments are again easily carried out. Finally, rather than deleting
the edges which meet and the associated junctions involved, the meeting
edges are simply re-connected as shown in Figure~\ref{fig: face-swap-illustration}.

\emph{Comments.}\,\, Note that the final configuration is similar
to the original configuration; in fact, with suitable facet motion,
the surface could return to its original configuration via another
face swap; the event is thus self-reversible in a sense. Also, since
both a facet pinch and a facet join occur simultaneously, the total
numbers of each surface element remain unchanged during this event.

\subsection{Vanishing Facets}

\label{sec: vanishing-facets}

\begin{figure}[ht]

\begin{centering}
\scalebox{.5}{\includegraphics{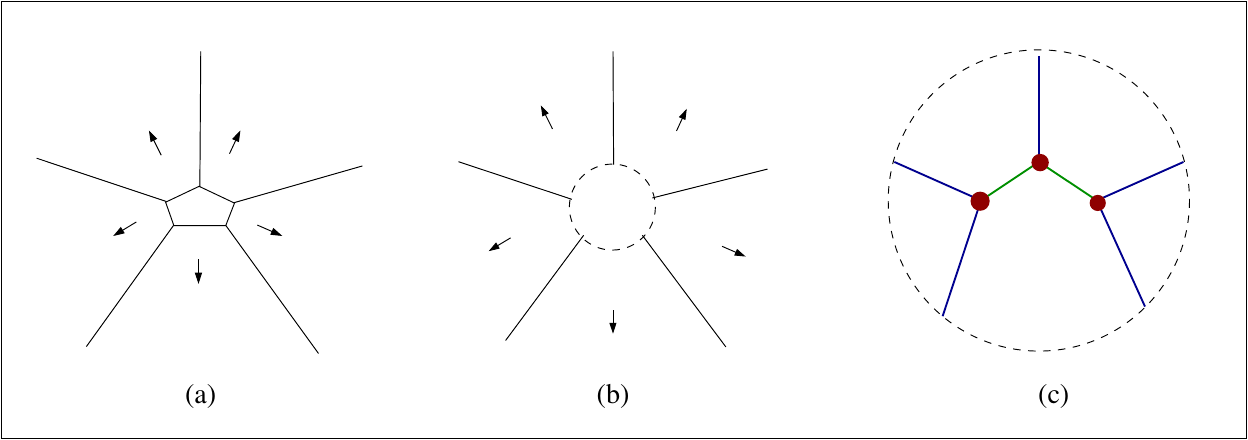}}
\caption[Example Vanishing Facet Event]{ (a) a single flat facet vanishing into a pentagonal well. (b) removing
the facet leaves a far field outside the dotted circle, which must
be reconnected. (c) the unique reconnection shown inside the dotted
circle. Arrows indicate gradients. }

\par\end{centering}

\label{fig: facet-removal-illustration} 
\end{figure}

The final class of topological event occurs when a facet shrinks to
zero area and is removed. However, as has been noted numerous times
previously in the context of cellular networks, very small facets
can result in stiff dynamics that are difficult to numerically simulate
accurately. For this reason, we follow previous authors by pre-emptively
removing facets with areas below some small threshold (but see Section~\ref{sec: pure-predictive-method}).
This process is summarized in Figure~\ref{fig: facet-removal-illustration}.
There, we see a single small flat facet vanishing into a pentagonal
well (\ref{fig: facet-removal-illustration}a). Being smaller than
the allowed threshold, it is removed, leaving a {}``hole'' in the
network (\ref{fig: facet-removal-illustration}b). The facets and
edges bordering this hole we call the \textbf{far field}, and they
need to be reconnected correctly to patch the hole. The correct reconnection
for this particular well is shown in Figure~\ref{fig: facet-removal-illustration}c.

The principal difficulty in this process occurs during the reconnection
step (Figure~\ref{fig: facet-removal-illustration}c). Here, we are
assigning new neighbor relationships to the far-field facets, which
also involves the creation of new edges and junctions to form boundaries
between them. In other cellular-network problems, these neighbor relationships
(and hence the reconnection) is usually chosen randomly, under the
reasoning that any error introduced is small enough to neglect and
quickly corrected%
\footnote{See, however, \cite{levitan-domany-1995,levitan-domany-1996}, where
the effects of this random choice in soap froths is investigated and
found to be significant. A deterministic method of re-connection is
proposed, based on the assumption that a cell loses sides as it shrinks
until it has only three. %
}. However, because the faceted-surface network represents a piecewise-planar
geometrical surface, we are not free to choose randomly. Since each
facet in the far field has a normal and a local height, neighbor relationships
determine junction locations and thus edge placement. However, the
final reconnection must be geometrically consistent - all facets must
be simply connected, and thus no edges may intersect. If we were to
randomly choose our neighbor relationships, the resulting reconnection
would likely fail this test, and would thus represent a non-physical
{}``surface.'' 

To guarantee a geometrically consistent reconnection, we must search
through all \emph{virtual} reconnections until we find one that does
not result in any self-intersecting facets. Several questions immediately
arise: 
\begin{enumerate}
\item How can we effectively characterize a {}``reconnection''? 
\item How many virtual reconnections are there to search? 
\item How can we efficiently list all these choices? 
\item Can we be sure a good reconnection exists? 
\item Is this reconnection unique? 
\end{enumerate}
For our method to be effective, all but the last of these questions
must be answered satisfactorily. The detailed answers to (1-3) are
found in the appendix, but we will summarize them here. The edges
and junctions created during an $O(n)$ reconnection may be effectively
characterized as a binary tree with $n-2$ nodes. The number of $m-$noded
binary trees is given by the \emph{Catalan number} $\mathcal{C}_{m}=\frac{2n!}{n!(n+1)!}$.
Finally, these trees may be efficiently listed using a greedy recursive
algorithm in $O(\mathcal{C}_{m})$ time. For the fourth question regarding
existence, we argue heuristically that a facet reaching zero area
proves the existence of its own reconnection, since a surface with
a zero-area facet is functionally the same as the surface with that
facet removed. We then assume the existence of that same reconnection
for some window of time before the facet reaches zero. A fuller proof
would appeal to manifold theory. Finally, the fifth question regarding
uniqueness is addressed in Section~\ref{sec: non-uniqueness}.

Having established these facts, we have a robust method for reconnecting
an arbitrary far field of facets. Before considering some special
cases of this method, let us summarize the general process so far:
Whenever facets smaller than a threshold area are detected, we:
\begin{enumerate}
\item remove them, leaving a hole in the mesh. 
\item list all virtual reconnections (VR's) as n-node binary trees. 
\item use associated neighbor relationships to find edge locations. 
\item test each VR until one with no intersecting edges is found. 
\end{enumerate}
We note that this approach represents a comprehensive reconnection
method for \emph{any} cellular network problem. Though it is necessary
for the faceted surface problem, it may be useful in any situation
where a verifiably optimal reconnection is sought.

\subsubsection{Special Case: Facets Disappearing in Groups}

\label{sec: volume-removal} 
\begin{figure}[ht]

\begin{centering}
\scalebox{.5}{\includegraphics{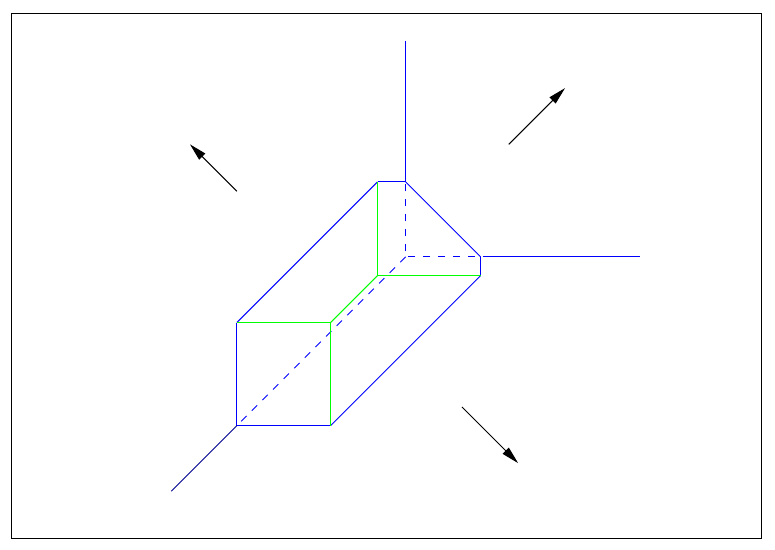}} \caption[Facets Vanishing in a group]{ Example of a group of disappearing facets. Reconstruction shown
in dotted lines. }

\par\end{centering}

\label{fig: volume-removal} 
\end{figure}

It is possible for \emph{groups} of facets
to shrink together, in such a way that they cannot be removed sequentially.
For an example, consider the configurations in Figure~\ref{fig: volume-removal}.
In such a case, it is necessary to identify and collect a contiguous
group of small facets for simultaneous deletion -- we call this a
\textbf{near field}. Any facet neighboring the near field is assigned
to the far field, which may be reconnected as described previously
after the near field is deleted.

To gather the near-field facets, we maintain a second, more liberal
threshold. Whenever a face shrinks below the first threshold, as described
above, its neighbors are recursively examined to collect those smaller
than the second threshold. This method is rather simplistic, and,
in cases of oddly-shaped pyramids, may not return the entire near
field. This, in turn, will result in an incorrect far field, which
will most likely be non-reconnectable. However, a group of facets
vanishing together eventually all head to zero area, and for some
window of time before they would physically vanish, all are small
enough to be detected in this way. Thus, we allow the code to {}``skip
over'' small facet combinations that it cannot remove successfully,
and try again during a future timestep.

\subsubsection{Special Case: Facets Disappearing as Steps}

\label{sec: step-removal} 
\begin{figure}[ht]

\begin{centering}
\scalebox{.5}{\includegraphics{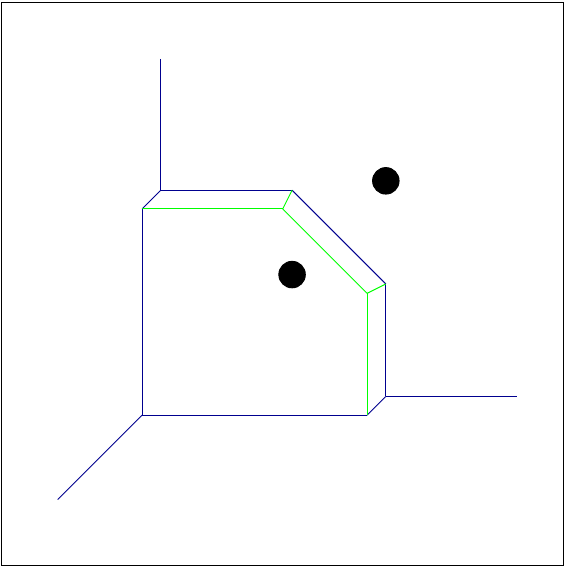}}
\scalebox{.5}{\includegraphics{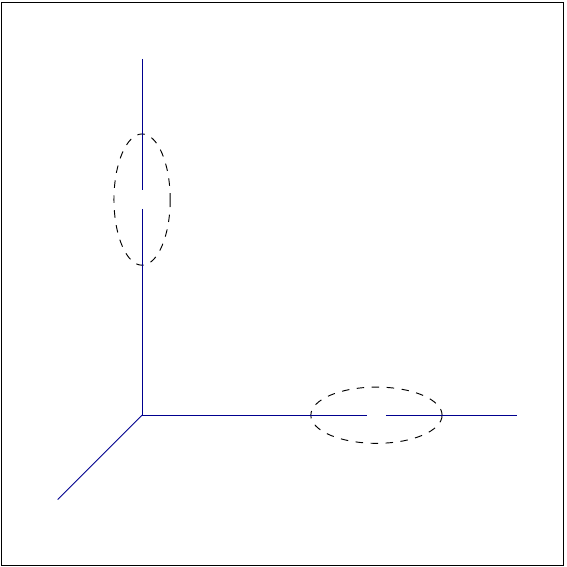}}
\caption[Facets Vanishing as a step]{ Example of a step removal. Left: A chain of small facets separates
two large facets of identical orientation. Right: The small facets
have been removed, and the large facets joined. }

\par\end{centering}

\label{fig: step-removal} 
\end{figure}

It is also possible, on high-symmetry
crystal surfaces, that the small facet or group of facets forms a
{}``step'' between two much larger facets of identical orientation,
but different height. Figure~\ref{fig: step-removal} illustrates
this situation, in which the near field is bounded by exactly four
facets, two of which have identical orientations. In such a case,
the final fate of the surface is that the small facets vanish \emph{as
the large facets join together}. The method described above contains
no provision for joining far-field facets during reconnection, and
so there is no way to reconnect the far field produced in this case.

Having identified a near field as forming a step, one solution is
to delete the small facets, then move the two large faces to the same
height and join them. This results in two pairs of unconnected edges,
which are each deleted and replaced with an appropriate single edge.
Since facet groups forming steps are, in fact, bordered by four facets
generically, a separate \noun{Repairman} class could be written to
handle this case. However, the small adjustment to the positions of
the large facets can lead to subtle problems, as will be seen in Section~\ref{sec: discretization and topology}.
Therefore, a more robust if less elegant approach is to simply add
one of the large parallel facets to the (step-forming) near field;
a good choice is the one with fewer edges. Since the far field surrounding
this modified near field requires no joins, it can be repaired using
the FFR method. Again, failures are possible as described in the above
section, but resolution is always possible near enough to the time
the event would physically occur.

\section{Discretization and Performance of Topological Events}

\label{sec: discretization and topology}

We have now discussed the general kinematics of a PADS, and surveyed
all topological events which may occur as the surface evolves. Before
our treatment is complete, however, we must consider with care the
application of a time-stepping scheme. The accurate performance of
topological events under such a scheme is problematic because, while
events on a continuously evolving surface happen at precise times
($E_{i}$ at $t_{i}$), any time-stepping method invariably skips
over these times. This has three consequences, concerning detection,
consistency, and accuracy. After discussing them briefly, we will
present three possible timestepping methods which illustrate them
in more detail.

\textbf{Detection.} \,\, Because timestepping will always skip over
moments of topological change, we must abandon hope of simply finding
topological events ready to perform. Instead, we must either look
ahead before each timestep and anticipate when events will occur (a
\emph{predictive} method); or step before looking, and then by examining
the network infer where events should have occurred (a \emph{corrective}
method). Class A events can be easily be detected either way, while
class B events are easier to correct, and class C are easier to predict.

\textbf{Consistency.} \,\, Once the occurrence of an event has been
detected by either means, it must be performed in a way that preserves
geometric consistency -- i.e., the network always corresponds to a
physical surface $s=h(\boldsymbol{x})$. For example, two joining
facets can only be mechanically fused if they exhibit the same local
height. If, in addition to the occurrence of an event, a detection
scheme can determine the exact time at which it occurred, then one
strategy is to move the network to the precise event time, at which
resolution is trivial. However, one may wish to attempt resolutions
at other times, and the geometric consequences of doing so must be
weighed.

\textbf{Accuracy} \,\, Finally, we must consider the possibility
of error that is produced during topological change. This error is
most easily understood if we view the evolving surface in its abstract
form as a highly nonlinear system of ODE's. The (usually configurational)
evolution function is moderated by the topological state; thus, topological
events can represent sudden, qualitative changes in the evolution
function. A naive time-stepping scheme which steps over these without
appropriate measures will produce large localized errors at moments
of topological change.

\subsection{Method 1. Predict Events, Travel Exactly to Each Event}

\label{sec: pure-predictive-method} Assume that, at all times, we
accurately predict the time and location of the next topological event%
\footnote{ An example of this approach may be found in the early soap froth
simulations of \cite{weaire-kermode-1983}, where edges and cells
shrinking to zero are anticipated. A similar predictive approach could
be developed for facets which become non-simply connected, by anticipating
the possibility of junctions crossing edges. %
}. Then, a straightforward timestepping strategy which avoids consistency
and accuracy concerns is to continually calculate the time of the
next topological event (accurate to the the order of the time-stepping
method), and then step from event to event. Under this approach, time
is divided into slices with constant equations of motion, guaranteeing
that that the system always evolves under the correct equations, and
accurately representing the continuously evolving surface. In addition,
high-order single-step methods such as Runga-Kutta methods may be
used to obtain high accuracy.

Though neatly eliminating consistency and accuracy concerns, this
method has a serious disadvantage. The frequency of topological events
scales with the system size, and since we can never step farther than
the next event, we effectively make the timestep dependent on system
size -- -- $\Delta t\sim\mathcal{O}\left(N^{-1}\right)$. Since moving
the system through a single timestep is itself an $\mathcal{O}\left(N\right)$
operation, then advancing the system through any $\mathcal{O}\left(1\right)$
period of time takes $\mathcal{O}\left(N^{2}\right)$ time. While
acceptable for the detailed study of a small surface, it is obviously
undesirable for the statistical study of large surfaces. This is chiefly
because, consistency concerns aside, it makes little sense to halt
the entire surface at every single topological event, when each of
these involves only a few facets. Thus, our next method has as its
chief objective the use of timesteps which are independent of system
size.

\subsection{Method 2. Use Fixed timestep -- Late Correction of Observed Events}

\label{sec: late-timestep-method-1} A second strategy is to take
fixed timesteps, use a corrective method of topological detection,
and attempt to perform topological corrections late. Since timestep
is independent of system size, many events will now occur per timestep,
the size of which is chosen to produce a fixed small \emph{percentage}
of facets undergoing topological change each step. While this approach
theoretically eliminates the $\mathcal{O}\left(N^{2}\right)$ contribution
to running time, it introduces hurdles to event detection, as well
as geometrically consistent and accurate resolution.

\textbf{Detection.} We just stated that, in this corrective detection
scheme, more than one event occurs per timestep. Whether or not this
is a problem depends on the \textbf{Domain of Influence} of each event,
defined to be the set of network elements that event affects. If these
sets contain no common elements, then the associated events occur
too far apart in space to affect each other -- they are independent.
Consequently, a detection routine can hand them in arbitrary order
to the repair routines, there to be confidently performed in isolation.
However, occasionally two or more domains of influence overlap. In
this situation, called a \textbf{Discrete Compound Event}, the associated
topological events are no longer independent, and a detection routine
can no longer ensure \emph{a priori} their correct, consistent resolution
when handed off. Even worse, the very signatures used to identify
separate events may be obscured in the resulting {}``tangle,'' such
that the routine does not even recognize what has happened. Given
the variety of event signatures described in Section~\ref{sec: topological events},
and the many combinations in which they might occur, creating a complete
list of all DCE's would be prohibitive if not impossible. Instead,
we reason that, on a random surface, no two events will ever occur
at exactly the same moment (It is possible to artificially construct
faceted surfaces such two or more events must occur simultaneously
-- we do not consider this case). Thus, if we simply refine our timestep
when necessary, formerly overlapping events can be sorted out, and
detected in sequence. A robust strategy for handling compound events
is thus to (a) \emph{retrace} the problematic timestep, (b) \emph{refine}
it into smaller slices, and (c) \emph{repeat} steps (a) and (b) recursively,
until only single events are detected.

\textbf{Consistency.} Since the surface is allowed to evolve unrepaired
past numerous topological events per timestep, surface regions near
these events will be geometrically inconsistent after the step. To
say the same thing, facets involved in the bypassed events will have
incorrect neighbor relationships. However, we have already classified
all possible events, so having identified which event occurred, and
which facets were involved, we know \emph{a priori} what the correct
neighbor relationships should be after the event. This knowledge,
along with knowledge of the position of each facet involved, allows
us to reconstruct the consistent surface that should have emerged
during the event%
\footnote{ This strategy is similar to the Far-Field Reconnection algorithm
described above, except that except that the correct neighbor relationships
are already known. However, FFR is a general algorithm for \emph{finding}
correct relationships between neighbors. Thus, many of the above topological
events described above may be performed {}``lazily,'' by simply
identifying the involved facets, and applying FFR. %
}.

Unfortunately, not all events can be consistently corrected at a late
time in this way. In particular, Facet Joins and Joining Facet Pinches
involve the joining of two facets that meet each other at a single
local height. Since this condition exists for only a single instant,
such events cannot be performed in a geometrically consistent way
at any time other than the {}``correct'' one. To accommodate this
requirement while preserving a topology-independent timestep, we are
forced to manually adjust the height of the joining facets before
the event is performed. Besides the error induced by this strategy
(discussed next), this need illustrates a second problem that can
arise. In a \textbf{Repair-Induced Inconsistency}, the very act of
performing one event, because it is done late, triggers a second event
that was not detected originally. An example is when the just-described
height adjustment required for the delayed repair of a facet join
triggers, say, a neighbor-switching event. Since this newly-triggered
event was not originally detected, the system is left in an inconsistent
state after all repairs are made. An ad-hoc strategy to find such
RII's is infeasible for the same reason as is a complete listing of
all possible Discrete Compound Events (indeed, an RII may produce
a DCE, which rules out a simple multiple-rechecking strategy). Thus,
a similar retrace/refine/repeat strategy is required, with the added
requirement that all events performed prior to detecting the RII must
first be undone.

\textbf{Accuracy.} Finally, as alluded above, repairing topological
events \emph{after} they occur can introduce large isolated errors.
This can be due to the {}``fudging'' required for the delayed repair
of facet joins and swaps, but more generally is caused by facets involved
in (uncorrected) topological events having been evolved under the
wrong equations of motion for part of the relevant timestep. Consider
an event $E_{i}:t^{j}<t_{i}<t^{j+1}$, with domain of influence $D_{i}$.
Since topological events likely correspond to a change in the surface's
evolution equation, the facets in $D_{i}$ are moved using the wrong
equations for the time interval $[t_{i},t^{j+1}]$. Since the equations
guiding $D_{i}$ are wrong by as much as $\mathcal{O}\left(1\right)$
for a time of order $\mathcal{O}\left(\Delta t\right)$, facets in
$D_{i}$ may accumulate $O(\Delta t)$ location errors during the
timestep in which the event occurs. Since the quantity of topological
events does not depend on $\Delta t$, the method retains first order
accuracy globally. However, this error introduces a barrier to achieving
higher-order accuracy later on.

\subsection{Sketch of Method 3. Localized Adaptive Replay}

The previous method, alas, contains one subtle problem that keeps
it from being a true $\mathcal{O}\left(N\right)$ method. This problem
is that the frequency of DCEs and RIIs, though small, still scales
with the system size, and these necessitate timestep refinement. So
although the late method does not have to explicitly step according
to the $\mathcal{O}\left(N^{-1}\right)$ time between topological
events, yet to accurately detect and resolve those events it is still
implicitly driven by the refinement strategy to step along a time
associated with DCEs and RIIs. While this characteristic time is longer
than that between individual topological events, and does not greatly
slow the simulation of tens of thousands of facets, it still results
in a method that is formally $\mathcal{O}\left(N^{2}\right)$, which
becomes prohibitive when considering systems of millions of facets.
Thus, we now sketch a third method, not yet implemented, which eliminates
this effect all together. In addition, the method allows us to perform
topological events in a way which confers all the accuracy benefits
of the first, predictive method.

We first re-state that, on any given timestep, \emph{most} facets
are not involved in any topological changes. While it was therefore
obviously wasteful to move the entire surface from event to event
in the first method, it is also conceptually wasteful to perform a
global retrace/refine/repeat step to DCEs and RIIs in the second method.
Instead, after every timestep, we should identify for each DCE/RII
the \textbf{Topological Subdomain} containing all facets involved
in the event. The few facets within these subdomains would be retraced/refined/repeated
as required, while the rest of the (unaffected) facets would be left
undisturbed in their post-timestep state. Since operating on a given,
constant number of facets takes $\mathcal{O}\left(1\right)$ time,
and since the number of events per timestep scales only like $\mathcal{O}\left(N\right)$,
we see that a single timestep and all associated corrections -- including
DCEs and RIIs -- can now be performed in $\mathcal{O}\left(N\right)$
time, with a final state that is guaranteed to be consistent. This
produces a true $\mathcal{O}\left(N\right)$ method. In addition,
this {}``Localized Replay'' strategy has an accuracy benefit. Regular,
recognized topological events also have easily identifiable topological
subdomains. If the facets within these domains are retraced, then
the predictive detection mechanism of the first method can be applied
within the domain to eliminate consistency and accuracy problems associated
with late removal.

One difficulty remains, however. Facets involved in topological events
may, under configurational facet-velocity laws, exhibit abrupt changes
in velocity a result of the event. During the remaining segment of
timestep, these facets may {}``break out'' of the subdomain initially
created to contain them, and begin interacting with facets outside
of it. Thus, we would need a mechanism to detect this, and start over
with a larger subdomain if it occurs. Finally, if subdomains can change
size, then there is the possibility that two nearby subdomains will
come to overlap as the algorithm progresses. Therefore, we must include
the ability to merge them if necessary, start over with the new, larger
sub-domain, and repeat adaptively until everything can be sorted.
This adaptivity ensures the robustness of the method, as highlighted
by the method's formal name of \textbf{Adaptive Localized Replay}.
The reader may note that the pattern of adaptive repetition is similar
to that used to resolve DCEs and RIIs above, and worry that another,
even smaller $\mathcal{O}\left(N^{2}\right)$ effect lurks in the
shadows. However, in both of the previous methods, such effects were
due to the \emph{global} response to a \emph{local} problem. Since
this latter method is designed to be localized, there is no longer
any mechanism to generate such effects.

\section{Demonstration and Discussion\label{sec: demonstration}}

We demonstrate our method using a sample dynamics associated with
the directional solidification of a strongly anisotropic dilute binary
alloy \cite{norris-davis-2007:dssa}. When a sample is solidified
at a pulling velocity which is greater than some critical value, solute
gradients caused by solute rejection at the interface create a solute
gradient which opposes and overcomes the thermal gradient, resulting
in a negative effective thermal gradient. In this environment, facets
move away from the freezing isotherm at a rate proportional to their
mean distance from the isotherm, as given by the dynamics 
\begin{equation}
V_{i}=\left<h\right>_{i}.\label{eqn: solidification-dynamics}
\end{equation}

\begin{sidewaysfigure}
\begin{centering}
\includegraphics[height=4cm]{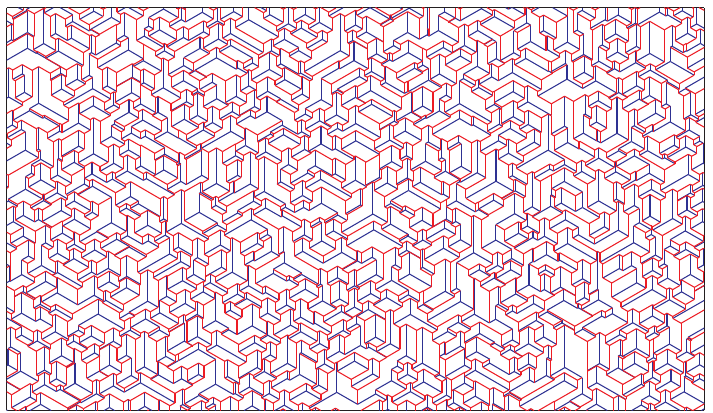}\includegraphics[height=4cm]{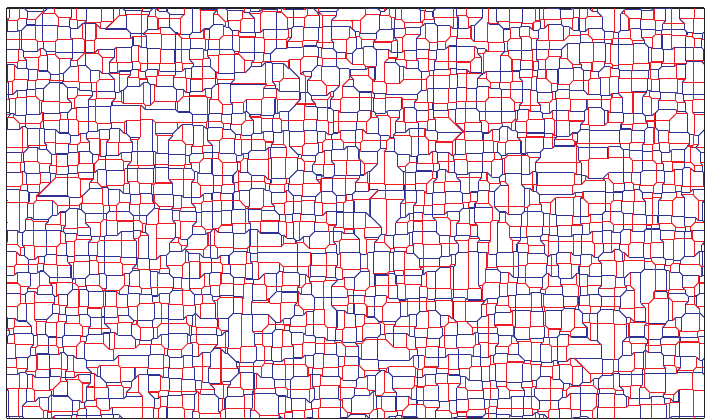}\includegraphics[height=4cm]{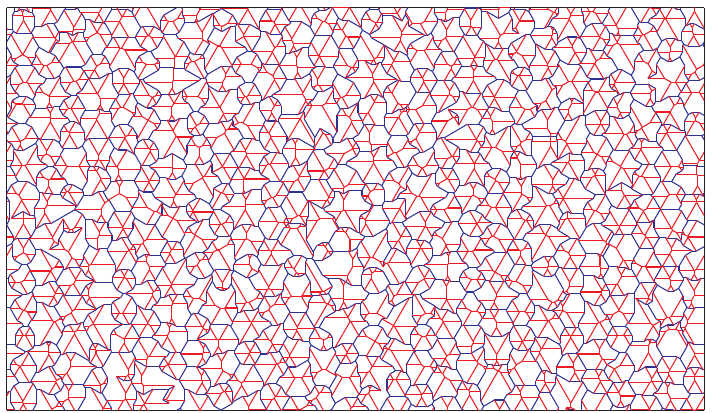}
\par\end{centering}

\begin{centering}
\includegraphics[height=4cm]{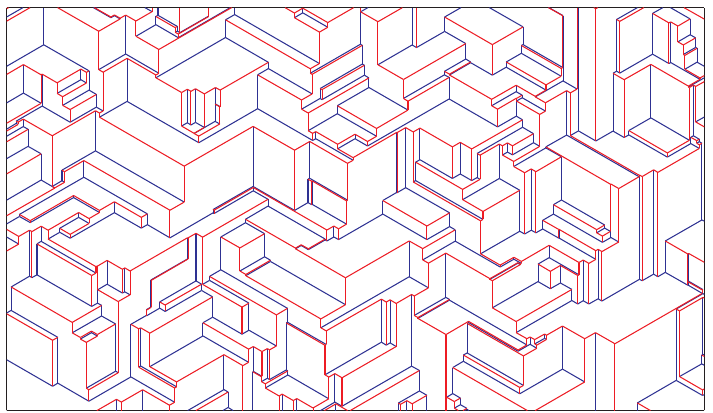}\includegraphics[height=4cm]{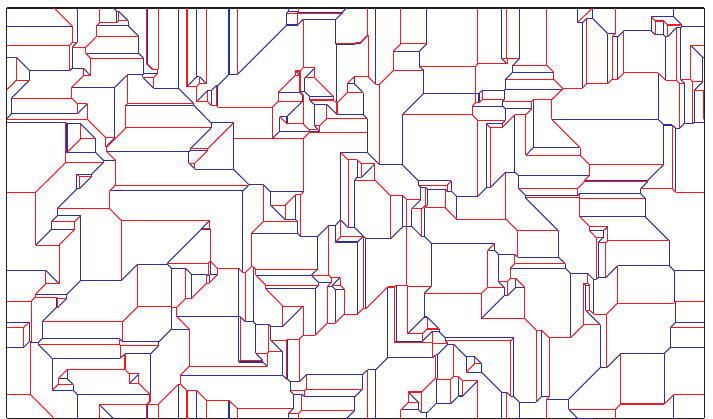}\includegraphics[height=4cm]{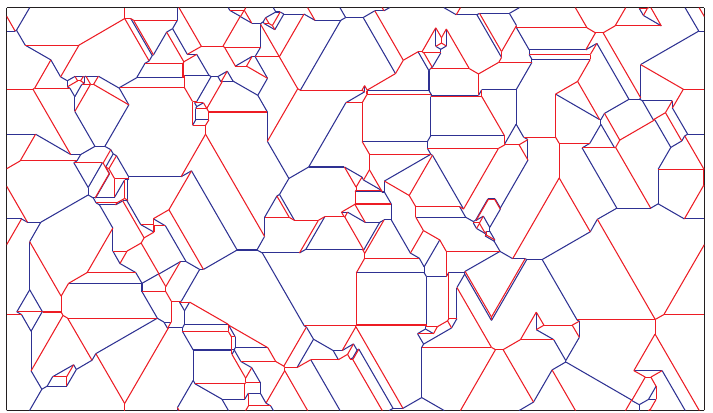}
\par\end{centering}

\centering{}\includegraphics[height=4cm]{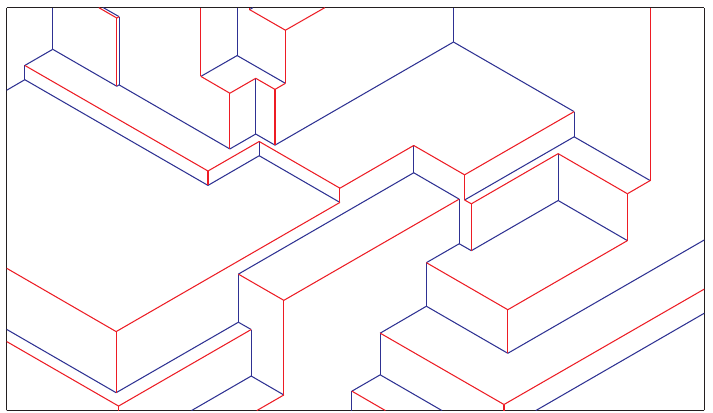}\includegraphics[height=4cm]{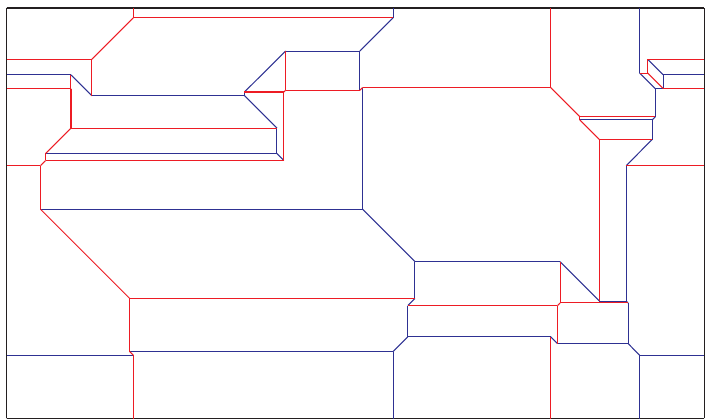}\includegraphics[height=4cm]{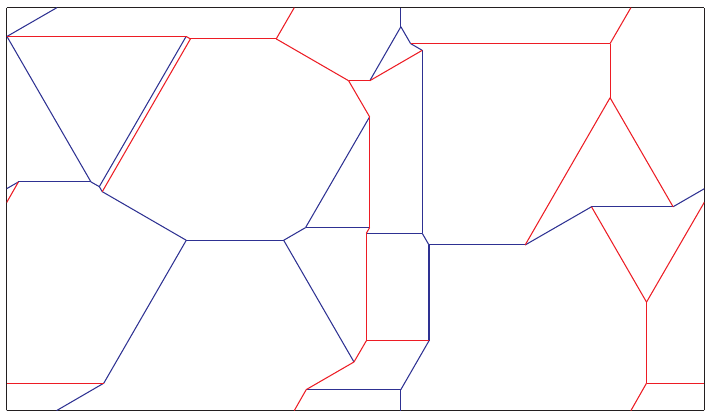}\label{fig: example-coarsening}\caption{Demonstration of the method. Here we present a top-down view of the
representative coarsening behavior for three small sample surfaces
with threefold symmetry (column 1), fourfold symmetry (column 2),
and sixfold symmetry (column 3). Spatial scale is constant, but irrelevant;
time increases down each column.}
\end{sidewaysfigure}

In Figure \ref{fig: example-coarsening}, this dynamics is applied
to surfaces with common three-, four-, and six-fold symmetries to
illustrate the flexibility of our topology-handling approach. A series
of snapshots from the coarsening surface are presented, in which surface
configurations near to many of the topological events described above
may be observed. (However, neither the Irregular Neighbor Switch,
Irregular Facet Pierce, nor Facet Pinch occur because no three facet
normals are coplanar in these symmetries; indeed, these events are
not expected to occur on most physical surfaces, and were included
for theoretical completeness.)%Following each set of snapshots, we illustrate the ease with which geometric data may be
%extracted by displaying the steady-state distributions of facet area and perimeter 
%associated with dynamic scaling.

About half of computational time is spent looking for topological
changes, which is significant but not prohibitive. With appropriate
timestep choice, using even the still-inefficient timestepping method
2 above, a surface of $25,000$ facets may be simulated to a $99$
percent coarsened state in around 10 minutes on currently available
workstations. With the implementation of method 3 above, this time
should be cut in half, and since method 3 is truly $\mathcal{O}\left(N\right)$,
a single million-facet simulation should take only a few hours. Looking
further ahead, since facet velocity calculations and topology checks
require only local information, the method should be easily parallelizable,
making possible even larger speed gains.

\section{Conclusions}

We have presented a complete method for the simulation of fully-faceted
interfaces of a single bulk crystal, with arbitrary symmetry, where
an effective facet velocity law is known. The surface, which is reminiscent
of two-dimensional cellular networks, is encoded numerically in a
geometric three-component structure consisting of facets, edges, and
junctions, and the neighbor relationships between them. Consistent
surface evolution specified by the facet velocity law is accomplished
via a simple relationship between facet motion and junction motion.
Although requiring the explicit handling of topological events, the
method is efficient, using the natural structure of the surface, and
accessible, allowing easy extraction of geometrical data. This combination
makes it ideal for the statistical study of extremely large surfaces
necessary for the investigation of dynamic scaling phenomena.

A comprehensive listing of all topological events has been presented.
These allow single-crystal surface with arbitrary symmetry (or no
symmetry at all) to be simulated. Events are classified into three
categories, representing three ways that surface elements can become
geometrically inconsistent. These are: edges which approach zero length,
facets which become constricted, and facets which approach zero area.
Resolution strategies for the former two classes can be determined
a priori, while repairing surface {}``holes'' left by vanishing
facets requires a novel Far-Field Reconnection algorithm, which iteratively
searches through all virtual reconnections to find one which produces
a consistent surface. Finally, intrinsic non-uniqueness of several
events is discussed; since ours is a purely kinematic method, decisions
regarding resolution of these events must be made ahead of time through
consideration of the dynamics or other physics.

In addition, a detailed discussion of the issues associated with a
discrete time-stepping scheme has been presented. The core issue is
that topological events, which occur at discrete times throughout
surface evolution, invariably fall between timesteps, with consequences
for the detection of events, as well as their geometrically consistent
and numerically accurate resolution. Since topological change corresponds
(under configurational facet velocity laws at least) to qualitative
changes in the local evolution function, some way to reach these in-between
times must be introduced, while recognizing that only a few facets
are involved in topological change during each timestep. A comparison
of three approaches showed that the optimal solution is one of Localized
Adaptive Replay, where large timesteps are taken to improve speed,
but local surface subdomains associated with topological change are
reverted, and then replayed in a way that re-visits events with the
necessary precision as necessary. While further work remains to implement
this approach, the method as presented is capable of comparing million-facet
datasets via averaged runs.\vspace{1in}

\textbf{Acknowledgements.} \,\, SAN was supported by NASA GSRP \#NGT5-50434.\bibliographystyle{unsrt}
\bibliography{tagged-bibliography}

\appendix

\section{Elaboration on Far-Field Reconnection}

\label{app: FFR details}

\begin{figure}[ht]

\begin{centering}
\scalebox{.65}{\includegraphics{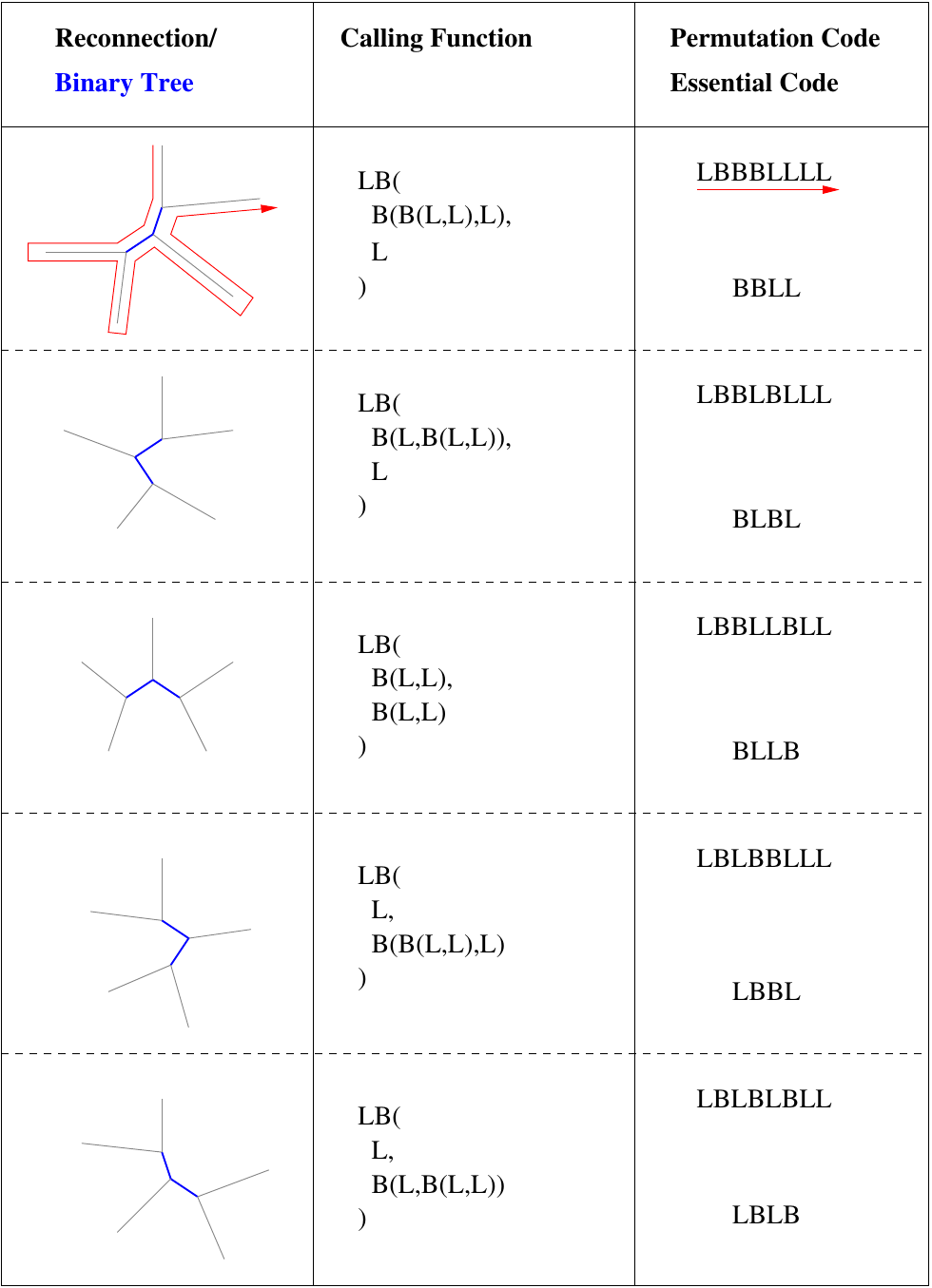}}
\caption[Method of listing binary trees]{ Our method of listing binary trees. The five possibilities for an
$O(5)$ far field are listed in the order given by our algorithm.
Left: The reconnection with {}``reconnection set'' binary tree in
blue. Middle: A recursive function describing the binary tree. Right:
The letter code and reduced letter code describing the binary tree.
The red lines in the top figure illustrate the {}``counter-clockwise
walk'' which also produces the letter code. A simple algorithm can
enumerate all abbreviated letter codes, which can be used to reconstruct
the binary trees for testing. }

\par\end{centering}

\label{fig: binary tree listing} 
\end{figure}

Our method of removing facets and facet groups requires patching a
{}``hole'' in the network left by the deleted facets. This requires
selecting a geometrically consistent reconnection from a list of potential,
or virtual reconnections. As outlined in the text, this involves searching
through a complete list of virtual reconnections and testing each
for geometric consistency. In this Appendix, we address in more detail
questions (1-3) posed in Section~\ref{sec: vanishing-facets} regarding
the details of this method. For convenience, we repeat them here: 
\begin{enumerate}
\item How can we effectively characterize a {}``reconnection''? 
\item How many potential reconnections are there to search? 
\item How can we efficiently list all potential choices? 
\end{enumerate}
We show here that an effective means of answer these questions is
to think of reconnections as extended binary trees. This characterization
enables us to easily count potential reconnections, distinguish them
through naming, and suggests an algorithm for efficiently listing
them for testing. An exhaustive illustration of the process is given
for the case of an $O(5)$ far field in Figure~\ref{fig: binary tree listing}.
It will be useful to refer to that diagram during the following discussion.

\subsection{Characterization: Binary Trees}

Patching network holes always involves finding unknown neighbor relations
between a given number of adjacent facets -- that is, no facets are
ever created, only edges and junctions. These are always connected
into a single graph. In fact, the edges and junctions created during
reconnection (the {}``reconnection set'') form a binary tree%  
\footnote{A binary tree is a graph consisting of edges and nodes such that:
(1) each node connects to 1,2 or 3 other nodes, (2) no cycles exist.
Condition (1) is met because we consider only triple junctions, while
condition (2) is met because the creation of new facets is excluded. %
}. In Figure~\ref{fig: binary tree listing}, the trees associated
with each possible reconnection are shown in thick blue lines. The
far-field edges touching the reconnection set shown in gray represent
the \emph{completion} of this tree. That is, they take the interior
tree, and add leaves to it so that every node of the interior tree
is a triple-node. Each virtual reconnection corresponds to a unique
interior tree and completed tree in this manner.

\subsection{Enumeration: The Catalan Number}

The counting of binary trees is, fortunately, a solved problem of
graph theory. Given $n$ nodes, they may be arranged in $\mathcal{C}_{n}$
distinct binary trees, where $\mathcal{C}_{n}$ is the $n^{\text{th}}$
\emph{Catalan Number}; 
\begin{equation}
\mathcal{C}_{n}=\frac{2n!}{n!(n+1)!}.
\end{equation}
 Now, re-connecting an $O(n)$ far field requires the creation of
$n-3$ edges and $n-2$ nodes; this may be visually confirmed for
the case $n=5$ in Figure~\ref{fig: binary tree listing}. This creates
an $n-2$ noded binary tree, and so an $O(n)$ far field has $\mathcal{C}_{n-2}$
virtual reconnections to search.

\subsection{Naming}

If take the completed binary tree and arbitrarily select a \emph{root
node}, then from each virtual interior tree we can generate a unique
sequence of letters which identify it and encode its construction.
This can be formed in one of two ways. The first way is to specify
the tree by a set of recursive function calls. Each branch point has
left and right branches, each of which may terminate in either a leaf,
or another branch point. The middle column of Figure~\ref{fig: binary tree listing}
gives such a function for each tree shown in the left column. The
second way is to walk around the tree in a counter-clockwise manner,
recording each branch point or leaf as it is encountered. Either method
produces a series of letters that uniquely identify the tree. Since
the beginning 'LB' and terminal 'LL' are guaranteed, we may use only
an abbreviated version consisting of $n-3$ of each letter.

\subsection{Listing}

The problem is now reduced to generating all possible letter combinations.
We can recursively build these combinations letter by letter using
a greedy algorithm which chooses 'B' over 'L' if possible. This approach
is subject to three restrictions which must be true of a {}``legal''
word. At each step, (a) $L\le B+1$, (b) $B\le n-3$, (c) $L\le n-3$.
The function we use is sufficiently short that we simply reproduce
it here: 
\begin{lyxcode}
function~treelist(n)\{

~~max~=~n-3

~~recursive\_treelist(max,0,0,'');

\}

function~recursive\_treelist(max,~leaves,~branches,~word)~\{

~~if~(branches~<~max)~\{~recursive\_treelist(max,~leaves,~branches+1,~word+'B')~\}

~~if~(leaves~<~max~and~leaves~<=~branches)~recursive\_treelist(max,~leaves+1,~branches,~word+'L')~;

~~if~(leaves~==~max~and~branches~==~max)~print(word)~;

\}
\end{lyxcode}
Thus, using this algorithm to generate a complete list of reconnection
labels, we perform the reconnection associated with each one, and
test it for geometric consistency. This allows us to efficiently find
the correct reconnection.

\section{Non-Uniqueness}

\label{sec: non-uniqueness}

We have already mentioned that certain topological events are ambiguous
in their resolution. Here we review the instances of non-uniqueness,
discuss their cause and implications, and suggest a method of treating
them in a numerical scheme.

\subsection{Review. Saddles.}

\begin{figure}[ht]

\begin{centering}
\scalebox{.5}{\includegraphics{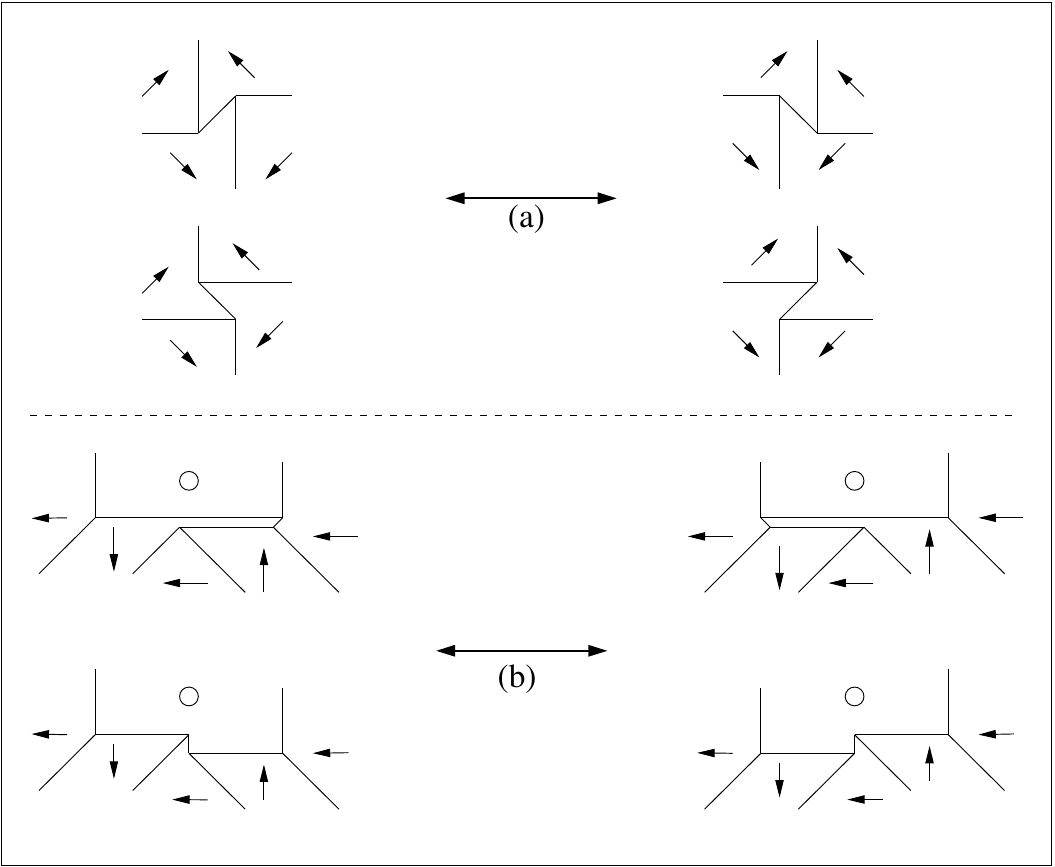}}
\caption[Saddle versions of Vanishing Edge events]{ Saddle versions of (a) the Neighbor Switch, and (b) the Irregular
Neighbor Switch. The saddle versions of these events produce additional
non-uniqueness beyond that described in the main text. }

\par\end{centering}

\label{fig: non-uniqueness} 
\end{figure}

We begin our list with an important ambiguity not discussed in the
main text. As noted by Thijssen \cite{thijssen-1995:diamond-simulation-2},
the lowly Neighbor Switch can be non-unique if the four facets involved
have a {}``saddle'' configuration -- where the edges neighboring
the vanishing edge (the emanating edges) form an alternating sequence
of two valleys and two ridges. This is illustrated in Figure~\ref{fig: non-uniqueness}a.
There, from either starting position on the left, two topological
resolutions are possible. One involves changing neighbor-relations,
while the other does not. This latter resolution, which we term a
{}``neutral pass,'' represents a Vanishing Edge event which needs
no resolution. However, it still results in a (prohibited) flipped
edge, so if it is to actually occur, a bookkeeping operation must
take place to correct this. Moving on, we recall the non-unique {}``gap
opener'' flavor of Irregular Neighbor Switch, which is not reproduced
here. However, as a Vanishing Edge event, it may also come in the
saddle variety, and admits a neutral-pass resolution option which
is displayed in Figure~\ref{fig: non-uniqueness}b. Finally, the
Irregular Facet Pierce event is also not reproduced.

% Resolution options for individual topological events

\subsection{Resolution Strategy: Analytical Backtracking}

None of these cases of ambiguity of resolution can be resolved at
the kinematic level. Since the particular evolution pathway leading
to such ambiguities is the product of a long chain of mathematical
reasoning, we therefore look backward in this chain to resolve them.
The kinematics are most directly provided by the dynamics, which is
a good first place to start. One option is to see which resolution
is more strongly self-reinforcing; if one choice under the given dynamics
immediately works to reverse itself, for example, the other option
should probably be chosen. Moving farther back the chain of reasoning,
one may consider the argument used to derive the given dynamics. For
example, the dynamics (\ref{eqn: solidification-dynamics}) comes
from the constant reduction of an undercooling energy by moving away
from the maximally-expensive $z=0$ isotherm. Therefore, an energy-informed
choice is to choose the resolution which minimizes this energy, i.e.,
maximizes the total integrated distance from $z=0$. Taking another
step back, perhaps such arguments come, as this one does, from a partial
differential equation describing surface evolution. Such equations
can be simulated directly, and the resulting resolution choices studied.
Ultimately, the original physical model may have to be considered,
perhaps at the atomic level. For each dynamics we wish to study, one
must apply this chain of reasoning to find the correct ambiguity-resolution
strategy. Ideally, considerations at all levels should produce identical
results.

\subsection{Implications for Far-Field Reconnections}

The non-uniqueness of particular topological events reflects a deeper,
underlying problem. For each non-unique event, facet heights of competing
resolutions are identical; only neighbor relationships between these
facets differ. Thus, non-uniqueness in topological events indicates
the presence of, and indeed is caused by, multiple possible FFR-style
reconnections of a given group of far-field facets. What are the implications
for the FFR algorithm of early facet removal? While it may seem helpful
at first to list prohibited resolution configurations and feed these
to the FFR algorithm, this approach grows increasingly brittle as
the number of prohibited configurations grows. Instead, a more robust
approach is to cause an FFR yielding multiple valid results to fail.
Having discovered resolution strategies for each individual event
as described above, we simply let these rules apply until the far
field is small enough to be unique (similar to \cite{levitan-domany-1995,levitan-domany-1996}).
A theory predicting when far fields will have multiple resolutions
would aid in this process.

\subsection{Comments}

Finally, we note that the benefits of phase-field and level-set methods,
which automatically capture topological change, are only clear if
topological changes are unique. These methods are ultimately only
kinematic, and offer no clear criteria for resolving kinematic non-uniqueness.
Thus, a dynamic selection criteria would seem to be required no matter
the method used, and indeed, the implicit handling of topology offered
by these methods may actually hinder the dynamic selection of kinematically
ambiguous events.

\end{document}